\documentclass[a4paper,11pt]{article}

\usepackage[utf8x]{inputenc}
\usepackage[T1]{fontenc}
\usepackage{color}
\usepackage{amssymb}
\usepackage{amsmath}
\usepackage{graphicx}
\usepackage{mathtools}
\usepackage{ragged2e}
\usepackage{physics}
\usepackage{dcolumn}% Align table columns on decimal point
\usepackage{bm}
\usepackage{graphicx}
\usepackage{braket}
\usepackage{verbatim} 
\usepackage[english]{babel}
\usepackage{hyperref}
\hypersetup{
citecolor=red,
colorlinks=true,
filecolor=red,
linkcolor=blue,
linktocpage=true,
urlcolor=blue
} 
\usepackage[titletoc,toc,title]{appendix}
\numberwithin{equation}{section}
\usepackage{cleveref}
\usepackage[margin=2.5cm]{geometry}
\usepackage{cite}
\usepackage{epsfig}
\usepackage{float}
\usepackage{amsfonts}
\usepackage{enumitem}
\usepackage[font={footnotesize,it}]{caption}

\numberwithin{equation}{section}

\begin{document}

\begin{center}
{\Large {\bf Holographic Entanglement Negativity and Thermodynamics in Backreacted AdS Black Hole}}

\vspace{8mm}

\renewcommand\thefootnote{\mbox{$\fnsymbol{footnote}$}}
Sanjay Pant,${}^{1}$\footnote{sanjaypant.phy@geu.ac.in}
Himanshu Parihar ${}^{2, 3}$\footnote{himansp@phys.ncts.ntu.edu.tw}, Pradeep Kumar Sharma ${}^{1}$\footnote{pradeepksharma.phd@geu.ac.in}

\vspace{4mm}
 \vskip 0.2cm
${}^{1}${\small \sl Department of Allied Sciences (Physics)}\\
%{\small \sl Physics division}\\
{\small \sl Graphic Era (Deemed to be University)}\\
{\small \sl  Dehradun, Uttarakhand 248002, India} 
     \vskip 0.2cm
${}^2${\small \sl Center for Theory-Computation-Data Science Research}\\
{\small \sl National Tsing-Hua University}\\
 {\small \sl Hsinchu 30013, Taiwan} 
 \vskip 0.2cm
${}^3${\small \sl Physics Division}\\
   {\small \sl  National Center for Theoretical Sciences}\\
     {\small \sl Taipei 10617, Taiwan} 
     \vskip 0.2cm
\end{center}

\vspace{6mm}
\numberwithin{equation}{section}
\setcounter{footnote}{0}
\renewcommand\thefootnote{\mbox{\arabic{footnote}}}

\begin{abstract}

We investigate holographic entanglement negativity (HEN) as a probe of mixed state quantum correlations in a deformed AdS black hole background with backreaction sourced by a string cloud. The bulk geometry is dual to a strongly coupled large-$N_c$ gauge theory at finite temperature, backreacted by a finite density of heavy static fundamental quarks. We analyze entanglement thermodynamics and establish a first law like relation in the small backreaction regime by identifying the deformed black hole at zero temperature as the natural ground state of the dual field theory. Using analytic expansions in the low and high temperature limits, we compute the HEN for adjacent, bipartite and disjoint subsystem configurations and examine its dependence on the backreaction parameter $\rho$. Our results reveal that at low temperature and high effective temperature, backreaction increases the HEN signaling enhancement of distillable quantum correlations. We further compare the behavior of HEN with holographic entanglement entropy (HEE) and mutual information (MI), demonstrating that entanglement negativity provides a sharper diagnostic of the competition between thermal and quantum correlations in holographic plasmas with matter backreaction.
\end{abstract}

\newpage
\tableofcontents
\newpage

\section{Introduction}\label{sec:intro}

Strongly coupled quantum systems appear in a wide range of physical settings ranging from condensed matter systems and cold atomic gases to the quark–gluon plasma formed in heavy-ion collisions. A central difficulty in analyzing such systems is the characterization of their quantum correlations, particularly in regimes where perturbative methods fail. The AdS/CFT correspondence \cite{Maldacena:1997re,Witten:1998qj,Gubser:1998bc} offers a powerful non-perturbative framework by relating strongly coupled gauge theories to weakly coupled gravitational theories in higher dimensional spacetimes. Within this holographic framework, quantum information theoretic measures such as entanglement entropy, mutual information, and computational complexity have emerged as valuable probes of the correlation structure of the dual field theories. Among these, the holographic computation of entanglement entropy through the Ryu–Takayanagi (RT) prescription and its covariant generalization \cite{Ryu:2006bv,Hubeny:2007xt} has played a central role.
The RT prescription has found extensive application in investigating the entanglement entropy of conformal field theories at finite temperature which are holographically dual to AdS-Schwarzschild black holes \cite{Cadoni:2010ztg, Solodukhin:2006xv, Hubeny:2012ry, Fischler:2012ca, Fischler:2012uv}. Detailed analyses of the resulting low and high temperature behaviors of holographic entanglement entropy led to the formulation of a thermodynamics like \textit{first law} for the entanglement entropy in \cite{Bhattacharya:2012mi, Allahbakhshi:2013rda}. Entanglement thermodynamics has also been investigated for charged black holes in AdS$_4$ backgrounds. In particular, the effects of temperature and charge on the entanglement entropy of the boundary field theory dual to charged AdS black holes were studied in \cite{Chaturvedi:2016kbk}, with extensions to higher-dimensional charged black holes presented in \cite{Karar:2018ecr}. Furthermore, entanglement thermodynamics has been explored in a variety of other scenarios including non-conformal and non-relativistic geometries \cite{Blanco:2013joa,Dey:2014voa,Mansoori:2015sit,Park:2015afa,Chakraborty:2014lfa,Mishra:2015cpa,Kundu:2016dyk,McCarthy:2017amh,OBannon:2016exv,Sun:2016til,Nadi:2019bqu,Saha:2019ado,Maulik:2020tzm}.

The entanglement entropy does not provide a faithful characterization of quantum correlations in mixed states as it captures both quantum and classical contributions. To extract the purely quantum correlations in such states, \emph{entanglement negativity} was introduced in quantum information theory \cite{Vidal:2002zz,Plenio:2005cwa}. Entanglement negativity serves as computable upper bound on distillable entanglement, and monotonic under local operations and classical communication (LOCC). Further it vanishes for separable states, thereby making it a robust measure of quantum correlations in mixed state scenarios. In recent years, entanglement negativity has gained prominence as a powerful diagnostic of quantum correlations in quantum field theories, capturing information inaccessible through entanglement entropy alone. A field-theoretic formulation based on the replica trick was developed in \cite{Calabrese:2012nk} enabling the computation of negativity in extended systems. This approach was later applied to finite temperature conformal field theories \cite{Calabrese:2014yza} and to lattice models such as the critical Ising chain \cite{Calabrese:2013mi}. The time-dependent behavior of entanglement negativity have also been explored in both local and global quantum quenches revealing universal features of entanglement spreading in conformal field theories \cite{Wen:2015qwa,Coser:2014gsa}. These results collectively establish entanglement negativity as a central tool in characterizing quantum correlations in many body systems. Motivated by these developments, several holographic prescriptions for entanglement negativity have been proposed \cite{Rangamani:2014ywa,Chaturvedi:2016rcn,Chaturvedi:2017znc,Jain:2017aqk,Jain:2017uhe,Malvimat:2018txq,Malvimat:2018ood,Chaturvedi:2016rft,Jain:2017xsu,Jain:2018bai,KumarBasak:2020viv,Mondal:2021kzj,Afrasiar:2021hld,Kudler-Flam:2018qjo,Kusuki:2019zsp,KumarBasak:2020eia,KumarBasak:2021lwm} extending the intuition gained from the RT prescription. 
The holographic proposals for entanglement negativity based on algebraic sums of bulk geodesics were further substantiated in \cite{KumarBasak:2020ams} for subsystems with spherical entangling surfaces through the analysis of replica symmetry breaking saddles in the gravitational path integral \cite{Dong:2021clv}. More recently, \cite{Dong:2024gud} has introduced a generalized framework for the holographic dual of negativity for general holographic states based on the modified cosmic brane proposal. The HEN turns out to be an effective measure to study the behavior of entanglement in the vicinity of critical points for strongly coupled gauge theory \cite{Karan:2023hfk}. These studies reveal that negativity is sensitive to critical behavior and captures features that are not visible through entanglement entropy alone. In parallel, HEN provides a robust characterization of quantum correlations in the presence of thermal and deformation effects in quantum field theories at finite temperature \cite{Pant:2024eno}. Together, these developments underscore the effectiveness of HEN as a better probe of mixed state entanglement in holographic frameworks. Closely related insights arise from the study of reflected entropy, that offers an alternative mixed state measure based on the entanglement structure of canonical purifications \cite{Dutta:2019gen}. While reflected entropy captures both quantum and classical correlations, its holographic dual via the entanglement wedge cross section provides complementary geometric information that enriches the understanding of correlation structure beyond what is accessible through negativity alone. Together, holographic entanglement negativity and reflected entropy describes a broad framework for probing mixed state entanglement in strongly coupled dual field theories.

Recent studies have highlighted the significance of investigating entanglement measures in bulk geometries that include matter backreaction as such effects capture additional physical features of the dual strongly coupled field theories. An instructive example is provided by a deformed AdS black hole geometry sourced by a homogeneous string cloud \cite{Chakrabortty:2011sp} which is holographically dual to a large-$N_c$ gauge theory at finite temperature backreacted by a uniform density of static heavy quarks. The backreaction parameter encodes the quark density and modifies both the near boundary and near horizon geometry, providing a controlled setup to investigate how heavy-flavor matter influences entanglement structure, thermodynamics and mixed-state quantum correlations. This setup has been extensively employed to explore transport properties, entanglement structure and chaotic behavior in the dual field theory \cite{Chakrabortty:2016xcb,Chakrabortty:2020ptb,Chakrabortty:2022kvq,Jain:2023xta}.

A natural question that arises is whether holographic entanglement entropy admits a relation similar to the first law of thermodynamics in such backreacted field theories. To address this, we study entanglement thermodynamics in deformed strongly coupled field theories in the context of AdS$_4$/CFT$_3$. Specifically, we analyze the dependence of the entanglement entropy on temperature and backreaction for a strip shaped subsystem in the boundary field theory dual to a deformed AdS$_4$ black hole. 
Furthermore, while the impact of backreaction on entanglement entropy and related probes has been widely investigated, the behavior of HEN in this background remains unexplored. Given that entanglement negativity provides a better probe of distillable quantum entanglement in mixed states, a systematic analysis in this context is both natural and timely. Therefore, in this work we present a detailed analysis of HEN for adjacent, bipartite and disjoint configurations in strongly coupled field theories dual to the string cloud deformed AdS black hole geometry. In our analysis we find that at low and high effective temperature regime, the HEN increases with backreaction in all three configurations at leading order. This observation supports the findings that extra degrees of freedoms introduced due to matter enhances the correlation in the strongly coupled field theory. Further, HEN for two disjoint subsystems allow us to compute the critical separation at which the distillable entanglement vanishes. In the leading order we observe that increasing backreaction increases the critical separation at which the negativity vanishes.

The structure of the paper is as follows. In \cref{sec:background}, we introduce the string cloud deformed AdS black hole geometry and its holographic dual interpretation. In \cref{sec:EErev} we provide a review of holographic entanglement entropy in this background, with particular emphasis on the analytic behavior in the low and high temperature regimes. In \cref{sec-en-th}, we analyze the holographic entanglement entropy in boundary field theories dual to deformed AdS$_4$ black holes at both zero and finite temperatures in the regime of small backreaction, and and establish a first law of entanglement thermodynamics at low temperatures. In \cref{sec:hen-adjacent}, we present the computation of holographic entanglement negativity for adjacent subsystems followed by the analysis of bipartite configuration in \cref{sec:HEN-bipartite}. The HEN for the configuration of two disjoint subsystems is examined in \cref{sec:hen-disjoint} followed by a critical separation analysis in \cref{sec:critical-separation}. Finally, in \cref{sec:summary}, we summarize our findings. 

\section{Deformed/Backreacted AdS Black Hole Spacetime}
\label{sec:background}

We begin by briefly reviewing the holographic setup corresponding to a strongly coupled gauge theory with a finite density of heavy static fundamental quarks, which is dual to a deformed AdS black hole geometry. This deformation arises due to the backreaction of a homogeneous distribution of long, non-interacting, and static strings stretched along the radial direction in the bulk spacetime. The endpoints of these strings are attached to the AdS boundary and represent static fundamental quarks, while the string bodies encode the associated gluonic degrees of freedom \cite{Chakrabortty:2011sp}.

The gravitational sector of the dual theory is described by the following $(d+1)$-dimensional Einstein-Hilbert action with a negative cosmological constant:
\begin{equation}
\label{action}
\mathcal{S}=\frac{1}{4\pi G_{d+1}}\int dx^{d+1}\sqrt{g}\left(\mathcal{R}-2\Lambda\right)+\mathcal{S}_{M},
\end{equation}
where the matter action $\mathcal{S}_M$ captures the effect of the string distribution and is given by
\begin{equation}
\label{matteraction}
\mathcal{S}_{M}=-\frac{1}{2}\sum_{i}\mathcal{T}_{i}\int d^2\xi \sqrt{-h}h^{\alpha\beta}\partial_{\alpha}X^{\mu}\partial_{\beta}X^{\nu}g_{\mu\nu}.
\end{equation}
Here, $g_{\mu\nu}$ denotes the $(d+1)$-dimensional spacetime metric, $h_{\alpha\beta}$ is the intrinsic worldsheet metric, and $\mathcal{T}_i$ represents the tension of the $i^\text{th}$ string. The corresponding Einstein’s equations, get modified by the energy-momentum tensor sourced by the strings:
\begin{equation}
\mathcal{R}_{\mu\nu}-\frac{1}{2}\mathcal{R} g_{\mu\nu}+\Lambda g_{\mu\nu}=8\pi G_N T_{\mu\nu},
\end{equation}
where the stress-energy tensor $T^{\mu\nu}$ associated with the string cloud is
\begin{equation}
T^{\mu\nu}=-\sum_{i}\mathcal{T}_{i}\int{d^2\xi\frac{1}{\sqrt{\mid{g_{\mu\nu}}}}\sqrt{-h^{\alpha\beta}}h^{\alpha\beta}\partial_{\alpha}X^{\mu}\partial_{\beta}X^{\nu}g_{\mu\nu}\delta_{i}^{d-1}(x-X_i)}.
\end{equation}
Considering all the strings are identical with tension $\mathcal{T}$ and adopt the static gauge choice $t=\xi^0$, $r=\xi^1$ gives the following number density of the strings
\begin{equation}
\label{metric}
\nonumber b(x)=\mathcal{T}\sum_{i=1}^{N}\delta_{i}^{(d-1)}(x-X_{i}),~~~\text{with}~ b>0,
\end{equation} 
where $N$ denotes the total number of strings. Assuming a uniform distribution over a spatial volume $V_{d-1}$, the average density becomes:
\begin{equation}
\label{metric2}
\nonumber \tilde{b}=\frac{1}{V_{d-1}}\int b(x)d^{d-1}x=\frac{\mathcal{T}N}{V_{d-1}}.
\end{equation}
 As suggested in \cite{Chakrabortty:2011sp}, the thermodynamic limit ($V_{d-1} \to \infty$), to keeps $N/V_{d-1}$ finite for maintaining constant quark density as $N\to \infty$.
The only non-vanishing components of the energy-momentum tensor are,
\begin{equation}
T_{00}=-\frac{\tilde{b}}{r^3}g_{tt}, \qquad T_{rr}=-\frac{\tilde{b}}{r^3}g_{rr}.
\end{equation}
The suitable ansatz for solving modified Einstein equation in AdS is 
\begin{equation}\label{mitr}
ds^2=-V(r)dt^2+\frac{dr^2}{V(r)}+\frac{r^2}{R^2}\delta_{ij}dx^{i}dx^{j},
\end{equation}
where the blackening factor $V(r)$ is defined as:
\begin{equation}
\label{V}
V(r)=K+\frac{r^2}{R^2}-\frac{2m}{r^{d-2}}-\frac{2b R^{d-3}}{(d-1)r^{d-3}}.
\end{equation}
Here $m$ is the mass parameter and $K$ denotes the curvature of the spatial boundary and takes values $-1$, $0$, or $1$ for hyperbolic, flat, or spherical geometries, respectively. In our analysis, we consider the planar ($K=0$) case only. Additionally, we define a dimensionless deformation parameter $b=\tilde{b}R^{d-1}$ for notational convenience.
Transforming to the inverse radial coordinate $z=R^2/r$, the metric takes the following familiar form:
\begin{equation}
\label{metric3}
ds^2=\frac{R^2}{z^2}\left(-h(z)dt^2+d\vec{x}^2+\frac{dz^2}{h(z)}\right),
\end{equation}
with the modified blackening function now expressed as
\begin{equation}
\label{h}
h(z)=1-\frac{2m}{R^{2d-2}}z^{d}-\frac{2b}{(d-1)R^{d-1}}z^{d-1}.
\end{equation}

It is convenient to rewrite $h(z)$ in terms of the horizon location $z_H$ and a dimensionless quark density parameter $\rho$:
\begin{equation}\label{hzr}
h(z)=\left[1-\rho\left(\frac{z}{z_{H}}\right)^{d-1}+(\rho-1)\left(\frac{z}{z_{H}}\right)^d\right],
\end{equation}
where
\begin{equation}
\rho =\frac{2bz_{H}^{d-1}}{(d-1)R^{d-1}}.
\end{equation}
Both $b$ and $\rho$ encode the information of the uniform string (quark) density in the bulk and on the boundary, respectively. These parameters will be used throughout the paper depending on the convenience. The Hawking temperature of this deformed AdS black hole is 
\begin{equation}
\label{tem}
T=-\frac{1}{4\pi}\frac{d}{dz}h(z)\Biggr|_{z=z_{H}}=\frac{(d-\rho)}{4\pi z_{H}}.
\end{equation}
Positivity of the temperature imposes the constraint $0\leq \rho \leq d$. The lower limit corresponds to the undeformed case ($b=0$), while the upper bound corresponds to the extremal limit with vanishing Hawking temperature.
The corresponding ADM mass of the black hole is related to the parameter $m$ via $M=\frac{(d-1)m V_{d-1}}{8\pi G_{d+1}}$, and the associated entropy density is:
\begin{equation}\label{entropy-density}
s=\frac{R^{d-1}}{4G_N^{d+1}}\frac{1}{z_{H}^{d-1}}.
\end{equation}
It is useful to introduce an effective temperature $T_f$, defined as:
\begin{equation}
\label{Tf1}
T_{f}=\frac{d}{4\pi z_{H}},
\end{equation}
which allows the entropy density to scale as $s \sim T_f^{d-1}$, a feature that becomes important in subsequent analyses of low temperature entanglement entropy.

The specific heat of the black hole computed from the variation of mass with temperature is given by
\begin{equation}
C = \frac{\partial M}{\partial T} =\frac{V_{d-1}(d-1)R^{2d-2}(d (d -1) R^{2d}z_H - 2 b R^4 z_H^{d})}{4 G_N^{d+1}  z_H^{d-1}(d (d-1) R^{2d}z_H + 
2 (d-2) b R^4 z_H^{d})}
\label{sh}
\end{equation}
 To ensure thermodynamic stability, the black hole must satisfy the condition:
\begin{equation}
z^{{\rm min}}_H = R\left(\frac{d(d-1)}{2 b}\right)^{\frac{1}{d-1}}.
\label{rmin}
\end{equation}
At zero temperature, the black hole acquires a negative minimum mass:
\begin{equation}
m^{{\rm min}} = -\frac{R^{d-1}}{d}\frac{b}{z^{{\rm min}}_H}.\end{equation}
 Above this bound, the specific heat remains positive and continuous, indicating that the black hole configuration is thermodynamically stable. For $b>0$, black holes with a negative mass parameter $m$ can exist within the range $m^{\rm min} \le m < 0$. In contrast, when $b \le 0$, black holes with negative $m$ do not possess real horizons. The corresponding free energy is given by:
\begin{equation}
{\cal F} 
=-\frac{(d-1)R^{2d} z_H + 2 b R^4 (d -2) z_H^d}{16 \pi z_H^{d+1}(d-1)}.
\label{f-e}
\end{equation}
 To examine the dynamical (geometric) stability of the background spacetime, one can carry out a gauge-invariant analysis of gravitational perturbations using the formalism developed in \cite{Ishibashi:2011ws, Kodama:2003jz, Ishibashi:2003ap, Kodama:2003kk, Kodama:2007ph,Lunin:2025yth}. Within this approach, the background spacetime is regarded as a product manifold:
\begin{equation}
M^{2+p} = {\mathcal{N}}^2 \times {\mathcal{K}}^p,
\end{equation}
Here, ${\mathcal{N}}^2$ denotes a two-dimensional spacetime with coordinates $(t,z)$, while ${\mathcal{K}}^p$ represents a $p$-dimensional maximally symmetric space spanned by the spatial coordinates $x^i$. Gravitational perturbations are classified into scalar, vector, and tensor modes on ${\mathcal{K}}^p$, which allows the Einstein equations to be analyzed in a decoupled manner. The tensor and vector sectors have been studied explicitly in \cite{Chakrabortty:2011sp}, where the background geometry was shown to be stable against these classes of perturbations.

\section{Revisiting the Holographic Entanglement Entropy and EWCS} \label{sec:EErev}
Since entanglement thermodynamics and holographic entanglement negativity are primarily studied in this work, it is natural to first review the computation of holographic entanglement entropy in the backreacted AdS black hole geometry, as introduced in \cref{sec:background}. Moreover, as reflected entropy is directly related to the entanglement wedge cross section, we also briefly review the EWCS. Both quantities were originally computed in \cite{Chakrabortty:2020ptb}.  

\subsection{Holographic Entanglement Entropy }
\label{sec:HEE}

We now revisit the computation of HEE for a single strip-like region in the backreacted AdS$_{d+1}$ black hole. This subsection provides all expressions later required for the computation of HEN.
According to the RT prescription, HEE is given by
\begin{equation}
S_A = \frac{\mathrm{Area}[\gamma_A]}{4 G_N} \,,
\end{equation}
where $\gamma_A$ is the extremal surface in the bulk anchored on the boundary of $A$.
Consider a rectangular strip like  subsystem $A$ with width $l$ along $x$ and infinite extent $L$ in other spatial directions defined by
\begin{equation}
x_1 \in \left[ -\frac{l}{2},\, \frac{l}{2} \right], \quad x_i \in \left[ -\frac{L}{2},\, \frac{L}{2} \right], \quad i=2,\dots,d-2\,.
\end{equation}
with $L\to\infty$. On a constant-time slice, the induced metric is
\begin{equation}
ds^2 = \frac{R^2}{z^2}\left[\frac{dz^2}{h(z)} + d{x_1}^{\,2}+d\vec{x_i}^{\,2}\right],
\label{metric-HEE}
\end{equation}
where the blackening function $h(x)$ is given by \cref{hzr}. The profile of the extremal surface $z(x_1)$ satisfies the RT area functional
\begin{equation}
\mathcal{A}[z] 
= R^{d-1}L^{d-2} 
\int_{-l/2}^{l/2} dx_1\, 
\frac{1}{z^{d-1}} 
\sqrt{1 + \frac{z'^{\,2}}{h(z)} }.
\label{area-functional}
\end{equation}
Due to translational symmetry in $x_1$ direction, the conserved Lagrangian leads to
\begin{equation}
\frac{1}{z^{d-1}}
\frac{1}{\sqrt{1+\frac{z'^2}{h(z)}}}
= \frac{1}{z_t^{d-1}},
\label{HEE-Hamiltonian}
\end{equation}
where $z_t$ is the turning point of the minimal surface. Solving for $z'(x_1)$ and integrating gives the width and turning-point relation
\begin{equation}\label{turning-point}
l/2=1/z_t^{d-1}\int_{0}^{z_t} z^{d-1}\left[h(z)\left(1-(z/z_t)^{2d-2}\right)\right]^{-1/2} dz.
\end{equation}
Substituting \eqref{HEE-Hamiltonian} back into the area functional gives the area of the extremal surface and consequently the HEE is given by the RT formula as
\begin{equation}\label{RT-area}
        S_A=\frac{
    L^{d-2}R^{d-1}}{2G^{d+1}_N} 
    \int_{0}^{z_t}\frac{dz}{z^{d-1}\sqrt{h(z)}\sqrt{1-\left(\frac{z}{z_t} \right)^{2d-2}}}.
\end{equation}

To obtain the HEE in terms of boundary parameters, it is necessary to express the turning point $z_t$ as a function of the strip width and temperature. However, evaluating the integral in \cref{turning-point} for arbitrary dimension $d$ and finite temperature is analytically intractable. As discussed in ~\cite{Chakrabortty:2020ptb}, we therefore analyze the problem in two regimes i.e. the low and high effective temperature limits where the integrals can be computed explicitly, and subsequently use these results to extract the corresponding behavior of the HEE.

\subsubsection*{Low temperature expansion:} 

By taking the low effective temperature limit i.e $T_f l \ll 1$ ( or $z_t/z_H \ll 1$) the expression of $z_t$ becomes
\begin{align}
z_t &= \frac{l}{2} \frac{\Gamma\left( \frac{1}{2d-2} \right)}{\sqrt{\pi}\,\Gamma\left( \frac{d}{2d-2} \right)} \Bigg[ 1 - \frac{\rho}{2d(2d-2)} \left( \frac{\Gamma\left( \frac{1}{2d-2} \right)}{\Gamma\left( \frac{d}{2d-2} \right)} \right)^2 \left( \frac{l}{2z_H} \frac{\sqrt{\pi}\,\Gamma\left( \frac{d}{2d-2} \right)}{\Gamma\left( \frac{1}{2d-2} \right)} \right)^{\!d-1} \nonumber\\
&\quad - \frac{1-\rho}{2(d+1)} \frac{\Gamma\left( \frac{2d}{2d-2} \right)\Gamma\left( \frac{1}{2d-2} \right)}{\Gamma\left( \frac{d}{2d-2} \right)\Gamma\left( \frac{1}{d-1} + \frac{1}{2} \right)} \left( \frac{l}{2z_H} \frac{\sqrt{\pi}\,\Gamma\left( \frac{d}{2d-2} \right)}{\Gamma\left( \frac{1}{2d-2} \right)} \right)^{\!d} + \cdots \Bigg].
\end{align}
Now the corresponding area function is
\begin{equation}\label{area-low-temp-ent}
\mathcal{A} = \frac{2}{d-2} \frac{L^{d-2} R^{d-1}}{\epsilon^{d-2}} + R^{d-1} \left( \frac{L}{l} \right)^{\!d-2} S_0 \left[ 1 + \rho S_2 \left( \frac{l}{z_H} \right)^{\!d-1} + (1-\rho) S_1 \left( \frac{l}{z_H} \right)^{\!d} + \cdots \right].
\end{equation}
Here $S_0,S_1$ and $S_2$ are only function of dimension parameter $d$, and are given by

\begin{equation}\begin{split}\label{Snotonetwo}
S_0&=\frac{2^{d-2}\pi^{\frac{d-1}{2}}\Gamma{\left(\frac{-d+2}{2d-2}\right)}}{(d-1)\Gamma{\left(\frac{1}{2d-2}\right)}}\left(\frac{\Gamma{\left(\frac{d}{2d-2}\right)}}
{\Gamma{\left(\frac{1}{2d-2}\right)}}\right)^{d-2}\\
S_1&=2^{-(d+1)}\pi^{-\frac{d}{2}}\frac{\Gamma{\left(\frac{1}{2d-2}\right)^{d+1}}}{\Gamma{\left(\frac{1}{d-1}+\frac{1}{2}\right)}\Gamma{\left(\frac{d}{2d-2}\right)^d}}
\left[\frac{\Gamma{\left(\frac{1}{d-1}\right)}}{\Gamma{\left(\frac{-d+2}{2d-2}\right)}}+\frac{(d-2)}{(d+1)}\frac{2^{\frac{1}{d-1}}}
{\sqrt{\pi}}\Gamma{\left(1+\frac{1}{2d-2}\right)}\right]\\
S_2&=2^{-d}\pi^{-(\frac{d-1}{2})}\left(\frac{\Gamma{\left(\frac{1}{2d-2}\right)}}{\Gamma{\left(\frac{d}{2d-2}\right)}}\right)^{d+1}\left[
\frac{\Gamma{\left(\frac{d}{2d-2}\right)}}
{\Gamma{\left(\frac{-d+2}{2d-2}\right)}}+\frac{(d-2)}{2d(d-1)}\right].
\end{split}
\end{equation}
These coefficients are negative for particular values of $d$. Finally, the HEE in low effective temperature limit is given by \cite{Chakrabortty:2020ptb}
\begin{equation}
S_A^{\mathrm{low}}
= \frac{R^{d-1}L^{d-2}}{4G_N^{d+1}}
\left[
\frac{2}{(d-2)\epsilon^{d-2}}
+ S_0\,l^{2-d}
+ S_0S_2\,\rho\,l
\left(\frac{4\pi T_f}{d}\right)^{d-1}
+ S_0S_1(1-\rho)\, l^{2}
\left(\frac{4\pi T_f}{d}\right)^d
+ \cdots
\right]
\label{HEE-lowT-final}
\end{equation}
The finite pieces $S_0\,l^{2-d}$ is negative contribution while $S_0S_2\,\rho\,l T_f^{d-1}$ is positive. On the other hand $S_0S_1(1-\rho)l^{2}T_f^d$ depends on the value of $\rho$. Therefore, the leading terms involving backreaction has positive contribution while the subleading term has reducing effect with respect to backreaction. Since we are in low temperature regime the dominating term with backreaction is $T_f^{d-1}$ that suggests enhancement of HEE with respect to backreaction. 

In \cref{HEE-lowT-final}, the first term is divergent containing UV-cut off ($\epsilon$). In the low-temperature regime, the entanglement entropy exhibits nontrivial modifications due to backreaction. Unlike the charged plasma case \cite{Kundu:2016dyk}, where the leading correction scales as $T_f^{\,d}$, here the dominant term is proportional to $T_f^{\,d-1}$. This distinction arises from the blackening function $h(z)$ near the boundary the $z^{\,d-1}$ contribution from the string density dominates over the mass term $\sim z^d$, making backreaction the leading source of corrections to the HEE at low effective temperature. It is also observed that this term follows a volume law scaling which suggests that the backreaction also introduces extensive classical correlations (thermal-like) into the subsystem.

\subsubsection*{High temperature expansion:} 

In the high effective temperature regime $T_f l \gg 1$ (or $z_t \to z_H$), the RT surface approaches the horizon. Although, the direct series expansions for $l$ and $A$ diverge in this limit, an appropriate combination of these quantities yields a finite and well behaved expression for the entanglement entropy as
\begin{equation}\label{subcombo}
\mathcal{A} - \frac{L^{d-2} R^{d-1}}{z_t^{\,d-1}}\,l = \frac{2 L^{d-2} R^{d-1}}{z_t^{\,d-2}} \int_0^1 \frac{\sqrt{h(u)}\,(1-u^{2d-2})du}{u^{\,d-1}\sqrt{1-u^{2d-2}}},
\end{equation}
where $u=\frac{z}{z_t}$. This leads to the HEE at high temperature limit as \cite{Chakrabortty:2020ptb}
\begin{equation}\label{hiee}
S^{high}_A = \frac{R^{d-1}}{4 G_N^{d+1}} \left[\frac{2}{(d-2)}\left(\frac{L}{\epsilon}\right)^{d-2}+ V \left( \frac{4\pi T_f}{d} \right)^{\!d-1} \left( 1 + 2 \left(\frac{d}{4\pi T_f l} \right)\,\widetilde{S}(\rho,d) \right) \right],
\end{equation}
where $V = l L^{d-2}$ is the volume of the strip subsystem $A$ in dual field theory and
\begin{equation}\label{stilde}
\widetilde{S}(\rho,d) = \frac{\sqrt{\pi}\,\Gamma\left( -\frac{d-2}{2d-2} \right)}{(2d-2)\,\Gamma\left( \frac{1}{2d-2} \right)} + \int_0^1 du \left[ \frac{\sqrt{1-u^{2d-2}}}{u^{\,d-1}\sqrt{h(z_H u)}} - \frac{1}{u^{\,d-1}\sqrt{1-u^{2d-2}}} \right].
\end{equation}
At high $T_f$, the leading term reproduces the thermal entropy density via its volume scaling with $\rho$ entering only at subleading order. Note that the quantity $\widetilde{S}(\rho,d)$ is independent of $l$ in the high temperature limit as $z_t \to z_H$ and depends only on the backreaction parameter $\rho$ and the dimension $d$.

Across both the low and high effective temperature limits, the quark density parameter $\rho$ plays a crucial role in determining the structure of the finite part of the entanglement entropy. At low $T_f$, the dominant correction scales as $T_f^{\,d-1}$ and volume of the subsystem which originates from the backreaction-induced $z^{\,d-1}$ contribution through $h(z)$. In contrast, at high $T_f$ the leading term coincides with the thermal entropy density while the dependence on $\rho$ manifests only through subleading corrections. 

\subsection{Entanglement wedge cross section and reflected entropy}
\label{sec:EWCS_SR}
Unlike HEE, the Entanglement Wedge Cross Section (EWCS) required two non-overlapping intervals (say $A_1$ and $A_2$). It is a geometric quantity associated with the minimal bulk surface that divides the entanglement wedge of two boundary sub regions $A_1$ and $A_2$ into two parts homologous to each of them. The EWCS has emerged as an important mixed state correlation measure in holography, conjectured to be dual to the entanglement of purification \cite{Takayanagi:2017knl, Nguyen:2017yqw}, odd entanglement entropy \cite{Tamaoka:2018ned}, entanglement negativity \cite{Kudler-Flam:2018qjo,Kusuki:2019zsp,KumarBasak:2020eia,KumarBasak:2021lwm} and widely studied at various occasions, see \cite{BabaeiVelni:2019pkw,Jokela:2019ebz,Jeong:2019xdr,Jain:2020rbb,Basu:2022nds,Jain:2022hxl,Jain:2022csf,Jiang:2024ijx}. In \cite{Dutta:2019gen} authors proposed a relation between \emph{reflected entropy} and EWCS via the following relation
\begin{equation}\label{eq:SR-prop}
S_R(A_1,A_2) = 2E_W(A_1,A_2).
\end{equation}

Now consider two identical parallel strip subsystems $A_1$ and $A_2$ on the boundary, each of width $l$ and separated by a distance $D$. On a constant time slice and at $x_1=0$, the induced metric reduces to
\begin{equation}
ds^2 = \frac{R^2}{z^2}\left[\frac{dz^2}{h(z)} + d\vec{x}^{\,2}_{d-2}\right].
\end{equation}
The EWCS between $A_1$ and $A_2$ is defined as the minimal area surface $\Gamma_{A_1\cup A_2}$ that connects the two RT surfaces $\gamma_{A_1}$ and $\gamma_{A_2}$ bounding the entanglement wedge of $A_1\cup A_2$. Its area is given by
\begin{equation}
E_W(A_1,A_2)= \frac{\mathcal{A}(\Gamma_{A_1 \cup A_2})}{4G_N^{d+1}}
= \frac{R^{d-1}L^{d-2}}{4G_N^{d+1}}
\int_{z_D}^{z_t}
\frac{dz}{z^{d-1}\sqrt{h(z)}},
\label{eq:EWCSdef}
\end{equation}
where $z_t$ and $z_D$ denote the turning points corresponding to the strip width $l$ and separation $D$, respectively. These turning points are determined by \cref{turning-point}. With an analogous expression for $D$ by replacing $z_t \to z_D$. Similar to the case of HEE, its analytic form is possible only in the low and high temperature limits.

\paragraph{Low-temperature EWCS:}
In the regime $T_f l \ll 1$, the turning points satisfy $z_t,z_D \ll z_H$, allowing a perturbative expansion of the lapse function $h(z)$. Substituting this into \eqref{eq:EWCSdef} yields
\begin{equation}
\begin{split}
E_W^{low}
&= \frac{R^{d-1}L^{d-2}}{4G_N^{d+1}}\Bigg[
K_0\!\left(\frac{1}{D^{d-2}} - \frac{1}{(D+2l)^{d-2}}\right)
+ K_1\,\rho\, l\!\left(\frac{4\pi T_f}{d}\right)^{\!d-1} \\
&\qquad\qquad\quad
-\, K_2(1-\rho)\,l(l+D)\!\left(\frac{4\pi T_f}{d}\right)^{\!d}
+ \mathcal{O}\big((T_f l)^{2d-2}\big)\Bigg],
\end{split}
\label{EWCSlow}
\end{equation}
where the coefficients $K_0,K_1,K_2$ depend only on the dimension $d$ and are given explicitly in \cite{Chakrabortty:2020ptb}. The first term corresponds to the vacuum contribution, while the $\rho$-dependent terms encode the leading and subleading backreaction effects. Notably, the correction proportional to $\rho T_f^{d-1}$ enhances the EWCS, indicating stronger mixed-state correlations in the presence of quark backreaction at low temperature.

\paragraph{High-temperature EWCS:}
In the opposite limit $T_f l \gg 1$, the turning points approach the horizon, $z_t,z_D \to z_H$, and the EWCS takes following the form
\begin{equation}
\begin{split}
E_W^{high}&=\frac{L^{d-2} R^{d-1}}{4G^{d+1}_{N}}T_{f}^{d-2}\Biggl\{-
   \left(\frac{2\sqrt{\pi}\Gamma \left(\frac{d}{2
   d-2}\right)}{\Gamma \left(\frac{1}{2
   d-2}\right)}\right)^{d-2}\left(\frac{1}{DT_{f}}\right)^{d-2}
   +\mathcal{C}\left(\frac{4\pi}{d}\right)^{d-2} \\
&\quad- \rho \left(\frac{d-2}{8 \sqrt{\pi } d(d-1) }\right)\!\left(\frac{\Gamma \left(\frac{1}{2 d-2}\right)}{\Gamma
  \left(\frac{d}{2 d-2}\right)}\right)^{3}\!\left(\frac{4 \pi }{d}\right)^{d-1}\!(DT_{f}) \\
&\quad-(1-\rho)\left(\frac{d-2}{8\pi(d+1)}\right)\!\frac{\Gamma
\left(\frac{2 d}{2 d-2}\right)}{\Gamma
\left(\frac{1}{2}+\frac{1}{d-1}\right)}\!\left(\frac{\Gamma \left(\frac{1}{2 d-2}\right)}{ \Gamma
\left(\frac{d}{2 d-2}\right)}\right)^{3}\!\left(\frac{4 \pi}{d}\right)^{d} (DT_{f})^2 \Biggr\},
\end{split}
\label{EWCShigh}
\end{equation}
with $\mathcal{C}$  is a function of $\rho$ and one can read form \cite{Chakrabortty:2020ptb}. In this regime, EWCS scales as $T_f^{d-2}$, and the backreaction enhances its magnitude monotonically. In the limit $\rho\to0$, both \eqref{EWCSlow} and \eqref{EWCShigh} reduce to the AdS--Schwarzschild results, providing a nontrivial consistency check.

\paragraph{Reflected entropy:}
In holographic theories, the reflected entropy $S_R(A_1,A_2)$ is proposed to be related to the EWCS via \eqref{eq:SR-prop} \cite{Dutta:2019gen}.
This relation is supported by explicit checks in static and time-dependent AdS$_3$/CFT$_2$ setups as well as tensor-network constructions \cite{Nguyen:2017yqw,Takayanagi:2017knl,Umemoto:2018jpc}. Consequently, the low and high temperature expressions for reflected entropy follow directly from \eqref{EWCSlow} and \eqref{EWCShigh} by an overall factor of two.

At low temperature, the leading backreaction correction $\propto \rho T_f^{d-1}$ enhances $S_R$, indicating that quark backreaction strengthens mixed state correlations beyond the vacuum contribution. At high temperature, $S_R$ inherits the thermal scaling $S_R\sim T_f^{d-2}$, with all backreaction dependent coefficients increasing monotonically with $\rho$. In the vanishing backreaction limit, $S_R$ reproduces the AdS-Schwarzschild results, confirming the robustness of the $S_R=2\,E_W$ prescription.

Although reflected entropy is geometrically determined by EWCS at the classical level, its interpretation as the entanglement entropy of a canonical purification provides additional insight beyond the purely geometric picture. In particular, while EWCS diagnoses the connectivity of the entanglement wedge, $S_R$ quantifies the total mixed-state entanglement shared between the subsystems, including correlations inaccessible to entanglement negativity. In the backreacted AdS black hole background, this distinction becomes important: EWCS and consequently $S_R$ is enhanced by backreaction at high temperature whereas HEN exhibits a qualitatively different behavior in the low-temperature regime. This contrast highlights the complementary roles of EWCS, reflected entropy and negativity in characterizing the entanglement structure of strongly coupled plasmas.

\section{Entanglement thermodynamics}\label{sec-en-th}

In this section, we restrict our analysis to $(3+1)$ dimensions (with $d=3$) for simplicity, although the results can be extended to higher dimensions. The metric of the deformed AdS$_4$ black hole with a planar horizon from \cref{metric3} takes the following form

\begin{equation}\label{metric-deformed-AdS4}
ds^2=\frac{R^2}{z^2}\left(-h(z)dt^2+d\vec{x}^2+\frac{dz^2}{h(z)}\right),
\end{equation}
with the blackening or lapse function expressed as
\begin{equation}\label{h-4d}
h(z)=1-\frac{b z^2}{R^2}-\frac{2 m\, z^3}{R^4}.
\end{equation}

The bulk theory is characterized by two primary parameters: the backreaction parameter $b$ and the black hole mass $m$ as described in \cref{sec:background}. These parameters are related via the horizon radius $r_H$ through the condition $h(r_H)=0$ yielding

\begin{equation}\label{mass-BH-T}
m=\frac{R^2 \left(R^2-b z_H^2\right)}{2 z_H^3}.
\end{equation}
The Hawking temperature for this deformed black hole can be expressed as 
\begin{equation}\label{temp-4d}
T=\frac{1}{4 \pi  z_H}\left(3-\frac{b z_H^2}{R^2}\right).
\end{equation}
By substituting the mass relation from \cref{mass-BH-T} into \cref{h-4d}, we can express the lapse function $h(z)$ solely in terms of the horizon radius $z_H$ and the backreaction parameter $b$ as follows

\begin{equation}
 h(z)=1-\frac{z^3}{z_H^3}+\frac{b z^2 (z-z_H)}{R^2 z_H}.
\end{equation}
This reparametrization reduces the effective parameters of the bulk theory to the backreaction $b$ and the horizon radius $z_H$. We then simplify the integrals for the HEE by defining $u = z/z_t$. Under this transformation, the expressions for the subsystem length $l$ and area $\mathcal{A}$ given in \cref{turning-point} and \cref{RT-area} respectively, leads to the following modified integral expressions

\begin{align}
\frac{l}{2} &= z_t \int_0^1\frac{u^2}{\sqrt{1-u^4}}\left( 1-\frac{u^3 z_t^3}{z_H^3}+\frac{b z_H^2}{R^2}\left(\frac{u^3 z_t^3}{z_H^3}-\frac{u^2 z_t^2}{z_H^2}\right)\right)^{-\frac{1}{2}}du, \label{l-4d}\\
{\cal A}&=\frac{2LR^{2}}{z_t}\int _0^1\frac{1 }{u^2\sqrt{1-u^4}}\left( 1-\frac{u^3 z_t^3}{z_H^3}+\frac{b z_H^2}{R^2}\left(\frac{u^3 z_t^3}{z_H^3}-\frac{u^2 z_t^2}{z_H^2}\right)\right)^{-\frac{1}{2}}du.\label{A-4d}
\end{align}
We utilize the subsystem length and area formulas given in \cref{l-4d} and \cref{A-4d} to investigate the thermal and backreaction effects on HEE. For a field theory dual to a deformed AdS$_4$ spacetime, the space of states is is characterized by both $T$ (via $z_H$) and $b$. This is in contrast to the Schwarzschild black hole in the AdS$_4$ bulk scenario where $T$ is the unique scale controlling the high and low temperature limits of the entropy. In contrast, for the deformed black hole, the backreaction parameter $b$ also plays a role, as is evident from the requirement of temperature positivity in the bulk which follows from \cref{temp-4d} as

\begin{equation}\label{zh-b-rel}
z_{H}\leq\frac{\sqrt{3}R}{\sqrt{b}}.
\end{equation}
In the relation above, the equality corresponds to the zero-temperature limit, while the inequality represents a black hole with finite temperature. According to the gauge/gravity duality, the field theory dual to a deformed AdS black hole occupies its ground state at zero temperature and enters an excited state at non-zero temperature. To derive a \textit{first law of entanglement thermodynamics}, we also analyze the HEE of the dual field theory at zero temperature. This situation bears resemblance to the case of extremal and non-extremal Reissner–Nordström black holes, however a key distinction is that in the present setup the black hole mass is negative at zero temperature as discussed in \cref{sec:background}, whereas it remains positive in the extremal case. We show that by defining this negative mass state as the reference ground state, the entanglement entropy satisfies a first law like behavior at low temperatures.

From the inequality in \cref{zh-b-rel}, one finds that the horizon radius $(z_H)$ is bounded from above by a quantity inversely proportional to the backreaction parameter $b$ of the deformed black hole. Consequently, the allowed range of the horizon radius is determined by the magnitude of the backreaction. When the backreaction is small, the horizon radius $(z_H)$ can take both large values, corresponding to the low temperature regime and small values, corresponding to the high temperature regime. In contrast, for zero-temperature deformed black holes, the equality condition in \cref{zh-b-rel} implies that the horizon radius $(z_H)$ is inversely related to the backreaction parameter $b$. As a result, a small backreaction leads to a large horizon radius in the zero-temperature case. In view of these observations, we examine the behavior of HEE in different regimes of the backreaction parameter for deformed AdS$_4$ black holes. Note that in \cref{sec:HEE} the HEE was analyzed at low and high effective temperatures while keeping the dimensionless backreaction parameter $\rho$ fixed. In the present section, however we focus on the low backreaction regime and study the HEE at both low and high temperatures, allowing us to explore its dependence on both parameters and to investigate the emergence of a first law like relation. We begin by analyzing the HEE of the subsystem $A$ in the boundary field theory dual to deformed black holes at zero temperature in the small backreaction limit. We then extend the analysis to deformed black holes at non-zero temperature in the high temperature regime, again restricting to small backreaction limit.

\subsection{Zero temperature}

In this section, we compute the HEE of subsystem $A$ in the boundary field theory dual to a deformed $\text{AdS}_4$ black hole at zero temperature. By imposing the condition $T_H = 0$ on the temperature expression in \cref{temp-4d}, we obtain the following constraint relating the backreaction parameter to the horizon radius as follows
\begin{equation}\label{b-zero-T}
b=\frac{3R^2}{z_H^2}.
\end{equation}
Thus, for deformed black holes at zero temperature, the small backreaction regime also corresponds to a large horizon radius $(l^2b\leq \frac{3R^2l^2}{z_H^2}<<1)$. With this condition, the lapse function from \cref{h-4d} simplifies
\begin{equation}
h(z)=1-\frac{3 z^2}{z_H^2}+\frac{2 z^3}{z_H^3}.
\end{equation}
Substituting this form of the lapse function into \cref{l-4d} and \cref{A-4d}, the integral expressions for the subsystem length $l$ and the extremal area $\mathcal{A}$ gives the following form
\begin{align}
 \frac{l}{2}&=z_t\int _0^1\frac{u^2 }{\sqrt{1-u^4}}\left(1-\frac{3 u^2 z_t^2}{z_H^2}+\frac{2 u^3 z_t^3}{z_H^3}\right)^{-\frac{1}{2}}du,\label{l-4d-zero}\\
 {\cal A}&=\frac{2LR^{2}}{z_t}\int _0^1\frac{1}{u^2\sqrt{1-u^4}}\left(1-\frac{3 u^2 z_t^2}{z_H^2}+\frac{2 u^3 z_t^3}{z_H^3}\right)^{-\frac{1}{2}}du. \label{A-4d-zero}
\end{align}
For large horizon radius $z_H$, the black hole lies deep in the bulk and is far from the extremal surface, so that $z_t<<z_H$. Consequently, we can expand $h(u)^{-1/2}$ as a power series in $z_t/z_H$ around $\frac{z_t}{z_H}=0$. Keeping terms up to the fourth order, the expansion is given by
\begin{equation}\label{h-appr-zero}
h(u)^{-\frac{1}{2}}\approx 1-\frac{u^3 z_t^3}{z_H^3}+\frac{3 u^2 z_t^2}{2 z_H^2}.
\end{equation}
Now solving the length integral in \cref{l-4d-zero} with the above \cref{h-appr-zero} perturbatively in terms of $(l/z_H)$ leads to the following expression for the turning point as
\begin{equation}\label{zt-zero}
    z_t=\frac{2 l \Gamma \left(\frac{5}{4}\right)}{\sqrt{\pi } \Gamma \left(\frac{3}{4}\right)}-\frac{16 l^3 \Gamma \left(\frac{5}{4}\right)^5}{\pi ^{3/2} \Gamma \left(\frac{3}{4}\right)^5 z_H^2}+\frac{8 l^4 \Gamma \left(\frac{5}{4}\right)^5}{\pi ^{3/2} \Gamma \left(\frac{3}{4}\right)^5 z_H^3}+\mathcal{O}(\left(\frac{z_t}{z_H}\right)^4).
\end{equation}

Using the same approximation for $h(u)^{-1/2}$ as in \cref{h-appr-zero}, we obtain an analytic expression for the extremal area by substituting it into \cref{A-4d-zero} yielding
\begin{equation}
    \begin{aligned}
{\cal A}&\approx \frac{2LR^{2}}{z_t}\int _0^1\frac{1}{u^2\sqrt{1-u^4}}\left(1-\frac{u^3 z_t^3}{z_H^3}+\frac{3 u^2 z_t^2}{2 z_H^2}\right)du \\
        &\approx \frac{2LR^{2}}{z_t}\left(\int _0^1\frac{1}{u^2\sqrt{1-u^4}}du-\int _0^1\frac{1}{u^2\sqrt{1-u^4}}\left(\frac{u^3 z_t^3}{z_H^3}-\frac{3 u^2 z_t^2}{2 z_H^2}\right)du\right).
    \end{aligned}
\end{equation}
The first term in the expression for $\mathcal{A}$ corresponds to the divergent contribution which is same as of pure AdS$_4$. We regularize the integral by imposing a UV cutoff $\epsilon = z_b/z_t$ and incorporate a counterterm $(-2LR^{2})/z_b$ to subtract the divergence. This procedure gives the following finite value for the area as
\begin{equation}\label{A-finite-zero}
    \begin{aligned}
 {\cal A}^{finite}&=  \frac{2LR^{2}}{z_t}\int _\frac{z_b}{z_{t}}^1\frac{1}{u^2\sqrt{1-u^4}}du-\frac{2LR^{2}}{z_{b}}+\frac{2LR^{2}}{z_t}\int _0^1\frac{1}{\sqrt{1-u^4}}\left(\frac{3z_t^2}{2 z_H^2}-\frac{u z_t^3}{z_H^3}\right)du\\
&= \frac{LR^{2}}{2z_t}\bigg[\frac{\sqrt{\pi } \Gamma (-\frac{1}{4})}{ \Gamma (\frac{1}{4})}+
 \frac{3 \sqrt{\pi } \Gamma \left(\frac{1}{4}\right) z_t^2}{2 \Gamma \left(\frac{3}{4}\right) z_H^2}-\frac{\pi  z_t^3}{z_H^3}\bigg].
    \end{aligned}
\end{equation}
After substituting the turning point $z_t$ from \cref{zt-zero} and performing a series expansion in $l/z_H$, the finite part of the extremal area (\cref{A-finite-zero}) leads to the following
\begin{equation}
  {\cal A}^{finite}=\frac{LR^{2}}{l}\left[\frac{\pi \Gamma \left(\frac{3}{4}\right) \Gamma \left(-\frac{1}{4}\right)}{\Gamma \left(\frac{1}{4}\right)^2}+\frac{\pi  \sqrt{2} l^2 \Gamma \left(\frac{5}{4}\right)}{\Gamma \left(\frac{3}{4}\right)^3 z_H^2}+\frac{16 \pi  \sqrt{2} l^3 \Gamma \left(\frac{5}{4}\right)}{\Gamma \left(-\frac{1}{4}\right)^3 z_H^3}+\mathcal{O}(\frac{l}{z_H})^4\right].
\end{equation}

The renormalized HEE of the subsystem $A$ in the boundary theory dual to the deformed black hole at zero temperature in the small backreaction regime can be obtained using the RT formula as
\begin{equation}\label{HEE-zero-T-4d}
    S_{ A}^{finite}=\frac{1}{4G_{N}^{4}}\frac{LR^{2}}{l}\left[\frac{\pi \Gamma \left(\frac{3}{4}\right) \Gamma \left(-\frac{1}{4}\right)}{\Gamma \left(\frac{1}{4}\right)^2}+\frac{\pi  \sqrt{2} \Gamma \left(\frac{5}{4}\right)b \,l^2}{3R^{2}\Gamma \left(\frac{3}{4}\right)^3 }+\frac{16 \pi  \sqrt{2} l^3 \Gamma \left(\frac{5}{4}\right)}{\Gamma \left(-\frac{1}{4}\right)^3 z_H^3}+\mathcal{O}(\frac{l}{z_H})^4\right].
\end{equation}
Therefore, in the zero-temperature case, the HEE receives corrections that are linear and quadratic in $l$ and $l^2$. Notably, the emergence of a term linear in $l$ distinguishes this geometry from the extremal RN black hole case \cite{Chaturvedi:2016kbk} where it is absent. By substituting the zero temperature mass $m_0=-R^{4}/z_H^3$ obtained from \cref{mass-BH-T} and \cref{b-zero-T} into above \cref{HEE-zero-T-4d}, the renormalized HEE can be expressed in the following compact form
\begin{equation}\label{EE-zero-temp}
 S_{ A}^{finite}\approx S_A^{AdS}+k_1bLl+k_2 m_0 Ll^2,
\end{equation}
where $S_A^{AdS}$ denotes the HEE for pure AdS$_4$ and constants $k_{1,2}$ are given by
\begin{equation}\label{k1-k2}
    k_1=\frac{1}{4G_{N}^{4}}\frac{\pi  \sqrt{2} \Gamma \left(\frac{5}{4}\right)}{3\Gamma \left(\frac{3}{4}\right)^3}, \quad k_2=\frac{1}{4G_N^4 R^{2}}\frac{ \Gamma \left(\frac{5}{4}\right)^2}{\Gamma \left(\frac{3}{4}\right)^2}.
\end{equation}
We see that in the small backreaction regime, the dominant contribution to the HEE of subsystem $A$ comes from the pure AdS$_4$ geometry. The subleading corrections in the \cref{EE-zero-temp} play a crucial role in subsequent sections where they form the basis for establishing a first law like relation for the HEE.

\subsection{Low temperature regime}

We now turn to the analysis of the subsystem $A$ in the boundary field theory dual to a deformed AdS$_4$ black hole with small backreaction at low temperature. As follows from \cref{zh-b-rel}, when both the temperature and the backreaction are small, the horizon radius $z_H$ becomes large. In this limit, one has $bz_H^2/R^2\sim 1$ and $z_H$ sufficiently large such that $ z_t \ll z_H $. Under these conditions, the lapse function $h(u)$ takes the form
\begin{equation}
 h(u)=1-\frac{u^3 z_t^3}{z_H^3}+\frac{b z_H^2 }{R^2}\left(\frac{u^3 z_t^3}{z_H^3}-\frac{u^2 z_t^2}{z_H^2}\right).
\end{equation}
By introducing the dimensionless parameter $\alpha = b z_H^2 / R^2$, we perform a Taylor expansion of $h(u)^{-1/2}$ around the $\frac{z_t}{z_H}=0 $. Neglecting higher order terms, we obtain the following approximated expression
\begin{equation}\label{h-low-approx}
h(u)^{-\frac{1}{2}}\approx 1+\frac{\alpha u^2 z_t^2}{2 z_H^2}+\frac{\left(1-\alpha\right)u^3 z_t^3}{2 z_H^3}.
\end{equation}
On utilizing the above expression in \cref{l-4d} for subsystem length $l$, we obtain
\begin{equation}\label{l-low-T}
 \frac{l}{2}\approx z_t\int_0^1 \frac{u^2 }{\sqrt{1-u^4}}\left(1+\frac{\alpha u^2 z_t^2}{2 z_H^2}+\frac{\left(1-\alpha\right)u^3 z_t^3}{2 z_H^3}\right)du.
\end{equation}
By inverting the above \cref{l-low-T}, we express $z_t$ as a function of the subsystem length $l$. Solving this relation perturbatively in the small parameter $(l/z_H)$ yields the following expression for the turning point as
\begin{equation}\label{zt-low}
z_t=\frac{2 l \Gamma \left(\frac{5}{4}\right)}{\sqrt{\pi } \Gamma \left(\frac{3}{4}\right)}-\frac{27 \alpha l^3 \Gamma \left(\frac{5}{4}\right)^5}{16 \pi ^{3/2} \Gamma \left(\frac{3}{4}\right) \Gamma \left(\frac{7}{4}\right)^4 z_H^2}+\frac{4 (\alpha-1) l^4 \Gamma \left(\frac{5}{4}\right)^5}{\pi ^{3/2} \Gamma \left(\frac{3}{4}\right)^5 z_H^3}+\mathcal{O}[(\frac{l}{z_H})^4].
\end{equation}

The extremal area is obtained by applying the approximation in \cref{h-low-approx} to the integral defined in \cref{A-4d} as follows
\begin{equation}\label{A-low-T}
 {\cal A}\approx \frac{2LR^{2}}{z_t}\int _0^1\frac{1}{u^2\sqrt{1-u^4}}\left(1+\frac{\alpha u^2 z_t^2}{2 z_H^2}+\frac{\left(1-\alpha\right)u^3 z_t^3}{2 z_H^3}\right)du.
\end{equation}
The expression for the extremal area in \cref{A-low-T} contains a divergent term identical to that of the pure AdS$_4$ vacuum. By applying the same regularization procedure used for the zero temperature case, we regularize the integral to isolate the finite contribution. The resulting finite part of the area is then given by
\begin{equation}\label{A-approx-low}
 {\cal A}^{finite}\approx L R^2 \left[\frac{\sqrt{\pi } \Gamma (-\frac{1}{4})}{ 2\Gamma (\frac{1}{4})z_t}+\frac{\sqrt{\pi } \alpha \Gamma \left(\frac{5}{4}\right)z_t}{\Gamma \left(\frac{3}{4}\right)z_H^2}+\frac{\pi  (\alpha-1) z_t^2}{4 z_H^3}\right].
\end{equation}
Now substituting $z_t$ from \cref{zt-low} into \cref{A-approx-low} and truncating the expansion at fourth order in $(l/z_H)$, the finite contribution to the extremal area becomes
\begin{equation}
    {\cal A}^{finite}=\frac{LR^{2}}{l}\left[\frac{\pi \Gamma \left(\frac{3}{4}\right) \Gamma \left(-\frac{1}{4}\right)}{\Gamma \left(\frac{1}{4}\right)^2}+\frac{9 \pi ^2 \alpha l^2}{128 \Gamma \left(\frac{3}{4}\right) \Gamma \left(\frac{7}{4}\right)^3 z_H^2}-\frac{8 (\alpha-1) l^3 \Gamma \left(\frac{5}{4}\right)^2}{\Gamma \left(-\frac{1}{4}\right)^2 z_H^3}+\mathcal{O}[(\frac{l}{z_H})^4\right].
\end{equation}
The holographic renormalized entanglement entropy of the subsystem $A$ for small backreaction at low temperature regime may now be obtained using the RT formula as follows
\begin{equation}
    S_A^{finite}=\frac{LR^{2}}{4 G_N^4\,l}\left[\frac{\pi \Gamma \left(\frac{3}{4}\right) \Gamma \left(-\frac{1}{4}\right)}{\Gamma \left(\frac{1}{4}\right)^2}+\frac{\sqrt{2} \pi\,\Gamma \left(\frac{5}{4}\right)b \,l^2}{3R^{2}\Gamma \left(\frac{3}{4}\right)^3 }+\frac{8 (1-\alpha) l^3 \Gamma \left(\frac{5}{4}\right)^2}{\Gamma \left(-\frac{1}{4}\right)^2 z_H^3}+\mathcal{O}[(\frac{l}{z_H})^4]\right].
\end{equation}
On using the expression for mass from \cref{mass-BH-T}, the above expression can be written in compact form as
\begin{equation}\label{EE-low-temp}
\begin{aligned}
 S_A^{finite}\approx S_A^{AdS}+k_1bLl+k_2 mLl^2,
 \end{aligned}
\end{equation}
where the constants $k_1$ and $k_2$ are same given in \cref{k1-k2}. We see that when backreaction and temperature both are small, the dominant contribution to the HEE for deformed black holes at non-zero temperature remains the pure AdS$_4$ similar to the case of zero temperature black holes.

\subsection{First law of entanglement thermodynamics}

We now examine entanglement thermodynamics within the small backreaction regime. According to the AdS/CFT correspondence, the deformed AdS black hole at zero temperature is identified as the holographic dual to the ground state of the boundary field theory. In contrast, the deformed black hole at finite temperature corresponds to an excited state. The change in HEE $\Delta S_A$ between the excited and ground states leads to a first law of entanglement thermodynamics as follows
\begin{equation}\label{first-law}
\Delta E_A=T_{\mathrm{ent}}\Delta S_A,
\end{equation}
where $T_{\mathrm{ent}}$ is known as the entanglement temperature and the change in HEE is computed by taking the difference between \cref{EE-low-temp} and \cref{EE-zero-temp} yielding
\begin{equation}\label{S-diff}
\Delta S_{A}=S_{A}-S_{A}^{T=0}=(m-m_0)k_2Ll^2.
\end{equation}
Furthermore, the energy increment associated with subsystem $A$ is defined as
\begin{equation}\label{E-diff}
\Delta E_{A}=\int_A dx^2 \  T_{tt}^{T\neq0}-\int_ A dx^2 \  T_{tt}^{T=0}=\frac{Ll}{4\pi G_N^4 R^2}(m-m_0),
\end{equation}
where $T_{tt}$ denotes the time component of the boundary stress-energy tensor and can be obtained using the holographic prescription of \cite{Balasubramanian:1999re}.
Now substituting the results from \cref{S-diff} and \cref{E-diff} into the first law like relation \cref{first-law}, the entanglement temperature is found to be
\begin{equation}
 T_{\mathrm{ent}}=\frac{\Delta E_A}{\Delta S_A}=\frac{1}{l}\frac{ 16\,\Gamma \left(\frac{3}{4}\right)^2}{\pi\,\Gamma \left(\frac{1}{4}\right)^2}.
\end{equation}
We observe that the entanglement temperature $T_{\mathrm{ent}}$ is proportional to the inverse of the subsystem size. This universal scaling behavior is consistent with the entanglement thermodynamics of the asymptotically AdS geometries as discussed in \cite{Bhattacharya:2012mi}. This suggests that the small backreaction does not alter the fundamental scaling of the first law and the HEE continues to satisfy a well defined thermodynamic relation in the perturbative regime.

\subsection{High temperature regime}

In this section, we analyze the high temperature behavior of the HEE for a subsystem $A$ when the bulk deformed black hole has a small backreaction and is in the high temperature regime. From \cref{zh-b-rel}, it follows that in the limit of high temperature ($l/z_H>>1$) and small backreaction, the parameter $bz_H^2/3R^2$ becomes much smaller than unity i.e $\frac{bz_H^2}{3R^2}\ll1$. By introducing the small parameter $\delta = bz_H^2/3R^2$, we can Taylor expand the lapse function $h(u)^{-1/2}$ around $\delta = 0$ yielding
\begin{equation}\label{h-approx-high-T}
   h(u)^{-\frac{1}{2}}\approx  \frac{1}{\sqrt{1-\frac{u^3 z_t^3}{z_H^3}}}+\frac{3 \frac{u^2 z_t^2}{z_H^2}\delta \left(1-\frac{u z_t}{z_H}\right)}{2 \left(1-\frac{u^3 z_t^3}{z_H^3}\right){}^{3/2}}.
\end{equation}
By employing the approximation for $h(u)$ given in \cref{h-approx-high-T}, the subsystem length $l$ and the extremal surface area $\mathcal{A}$ are evaluated using \cref{l-4d} and \cref{A-4d} respectively as follows
\begin{align}
    l&=2 z_t \int_0^1 \frac{u^2 }{\sqrt{1-u^4}}\left(\frac{1}{\sqrt{1-\frac{u^3 z_t^3}{z_H^3}}}+\frac{3 \delta  u^2 z_t^2 \left(1-\frac{u z_t}{z_H}\right)}{2 z_H^2 \left(1-\frac{u^3 z_t^3}{z_H^3}\right){}^{3/2}}\right) \, du\label{l-high-T-int}\\
    \mathcal{A}&=\frac{2LR^{2}}{z_t}\int_0^1 \frac{1}{u^2 \sqrt{1-u^4}} \left(\frac{1}{\sqrt{1-\frac{u^3 z_t^3}{z_H^3}}}+\frac{3 \delta  u^2 z_t^2 \left(1-\frac{u z_t}{z_H}\right)}{2 z_H^2 \left(1-\frac{u^3 z_t^3}{z_H^3}\right){}^{3/2}}\right) \, du\label{A-high-T-int}.
\end{align}
The integrals above do not admit simple closed forms. To proceed, we utilize the following identities:
\begin{equation}\label{identities}
     \frac{1}{\sqrt{1-x}}=\sum_{n=0}^\infty \frac{\Gamma(n+\frac{1}{2})}{\sqrt{\pi}\Gamma(n+1)} x^n,
  \frac{1}{{(1-x)}^{\frac{3}{2}}}=\sum_{n=0}^\infty \frac{2\Gamma(n+\frac{3}{2})}{\sqrt{\pi}\Gamma(n+1)} x^n, \int_0^1 \frac{u^m}{\sqrt{1-u^4}}du=\frac{\sqrt{\pi}}{4}\frac{\Gamma(\frac{m+1}{4})}{\Gamma(\frac{m+3}{4})},
\end{equation}
to express the integrals into a more tractable form. Consequently, the expression for the subsystem length given in \cref{l-high-T-int} can be rewritten as
\begin{equation}\label{l-series}
    \begin{aligned}
\frac{l}{2 z_t} =& \sum_{n=0}^\infty \frac{\Gamma(n+\frac{1}{2})}{\sqrt{\pi}\Gamma(n+1)}\left(\frac{z_t}{z_H}\right)^{3n}
 \int_0^1\frac{u^{3n+2}}{\sqrt{1-u^4}}+\delta \sum_{n=0}^\infty \frac{3\Gamma(n+\frac{3}{2})}{\sqrt{\pi}\Gamma(n+1)}\left(\frac{z_t}{z_H}\right)^{3n+2}
 \int_0^1\frac{u^{3n+4}(1-u\frac{z_t}{z_H})} {\sqrt{1-u^4}} \\
 \frac{l}{z_t} =& \sum_{n=0}^\infty \frac{\Gamma(n+\frac{1}{2})}{2\Gamma(n+1)}\frac{\Gamma(\frac{3n+3}{4})}{\Gamma(\frac{3n+5}{4})}\left(\frac{z_t}{z_H}\right)^{3n}
 +\frac{3\delta}{2}\sum_{n=0}^\infty \frac{\Gamma(n+\frac{3}{2})}{\Gamma(n+1)}\frac{\Gamma(\frac{3n+5}{4})}{\Gamma(\frac{3n+7}{4})}\left(\frac{z_t}{z_H}\right)^{3n+2} \\
 &\quad-\frac{3\delta}{2}\sum_{n=0}^\infty \frac{\Gamma(n+\frac{3}{2})}{\Gamma(n+1)}\frac{\Gamma(\frac{3n+6}{4})}{\Gamma(\frac{3n+8}{4})}\left(\frac{z_t}{z_H}\right)^{3n+3}.
    \end{aligned}
\end{equation}
Let us now examine the asymptotic behavior of the terms in the above expression to identify potential divergences. For large $n$, the first series scales as $\frac{1}{\sqrt{3}\,n}(z_t/z_H)^{3n}$, while the remaining two series behave as $ \frac{2}{\sqrt{3}}(z_t/z_H)^{3n} $. Upon isolating these terms, it becomes evident that the divergences in the latter two series cancel one another, leading to the following simplified expression as
\begin{equation}\label{l-series-high-T}
    \begin{aligned}
\frac{l}{z_t} =& \frac{\sqrt{\pi}}{2}\frac{\Gamma(\frac{3}{4})}{\Gamma(\frac{5}{4})}+ \sum_{n=1}^\infty\left( \frac{\Gamma(n+\frac{1}{2})}{2\Gamma(n+1)}\frac{\Gamma(\frac{3n+3}{4})}{\Gamma(\frac{3n+5}{4})}-\frac{1}{\sqrt{3}n}\right)\left(\frac{z_t}{z_H}\right)^{3n}\\
 &+\frac{3\delta}{2}\sum_{n=0}^\infty \left( \frac{\Gamma(n+\frac{3}{2})}{\Gamma(n+1)}\frac{\Gamma(\frac{3n+5}{4})}{\Gamma(\frac{3n+7}{4})}-\frac{2}{\sqrt{3}}\right)\left(\frac{z_t}{z_H}\right)^{3n+2} 
  -\frac{3\delta}{2}\sum_{n=0}^\infty\left( \frac{\Gamma(n+\frac{3}{2})}{\Gamma(n+1)}\frac{\Gamma(\frac{3n+6}{4})}{\Gamma(\frac{3n+8}{4})}-\frac{2}{\sqrt{3}}\right)\left(\frac{z_t}{z_H}\right)^{3n+3}\\
&-\frac{1}{\sqrt{3 }}\log\left[1-\left(\frac{z_t}{z_H}\right)^3\right]
+\sqrt{3}\delta \frac{(\frac{z_t}{z_H})^2}{(1+\frac{z_t}{z_H}+(\frac{z_t}{z_H})^2)}.    
    \end{aligned}
\end{equation}
As described in\cite{Hubeny:2012ry}, for a static asymptotically AdS black hole, the extremal surface cannot penetrate the event horizon, necessitating that $z_t < z_H$. In the high temperature limit, however, the turning point approaches the horizon such that $z_t \to z_H$. By introducing a small positive parameter $\epsilon \ll 1$, we can parameterize this proximity as $z_t = (1 - \epsilon)z_H$. Taking the limit $\epsilon \to 0$ in \cref{l-series-high-T}, the series converges to the following compact leading order expression as follows

\begin{equation}\label{l-exp}
 \frac{l}{z_H} = -\frac{1}{\sqrt{3}}\log[3\epsilon]+C_1+\delta C_2+\mathcal{O}[\epsilon] ,
\end{equation}
where the coefficients $C_1$ and $C_2$ are given by

\begin{equation}
    \begin{aligned}
 C_1 =& \frac{\sqrt{\pi}}{2}\frac{\Gamma(\frac{3}{4})}{\Gamma(\frac{5}{4})}+ \sum_{n=1}^\infty\bigg( \frac{\Gamma(n+\frac{1}{2})}{2\Gamma(n+1)}\frac{\Gamma(\frac{3n+3}{4})}{\Gamma(\frac{3n+5}{4})}-\frac{1}{\sqrt{3}n}\bigg)  \\
  C_2 =& \frac{1}{\sqrt{3}}+\frac{3}{2}\sum_{n=0}^\infty\bigg( \frac{\Gamma(n+\frac{3}{2})}{\Gamma(n+1)}\frac{\Gamma(\frac{3n+5}{4})}{\Gamma(\frac{3n+7}{4})}-\frac{2}{\sqrt{3}}\bigg)-\frac{3}{2}\sum_{n=0}^\infty \bigg( \frac{\Gamma(n+\frac{3}{2})}{\Gamma(n+1)}\frac{\Gamma(\frac{3n+6}{4})}{\Gamma(\frac{3n+8}{4})}-\frac{2}{\sqrt{3}}\bigg).
\end{aligned}
\end{equation}
On substituting the form of $z_H$ from \cref{temp-4d} in \cref{l-exp} leads to the following form of the parameter $\epsilon$ as
\begin{equation}
    \epsilon\approx\epsilon_0\,e^{-\frac{4\pi l T}{\sqrt{3}}(1+\delta)},
\end{equation}
where $\epsilon_0=\frac{e^{\sqrt{3}(C_1+\delta C_2)}}{3}$.

In evaluating the area integral given in \cref{A-high-T-int}, divergent part is confined to the first term. Following \cite{Fischler:2012ca}, we isolate the divergence and express the remaining finite contribution as
\begin{equation}
{\cal A}^{finite}=\frac{2LR^{2}}{z_t}\int_{\frac{z_b}{z_t}}^1 \frac{1}{u^2 \sqrt{1-u^4}} \left(\frac{1}{\sqrt{1-\frac{u^3 z_t^3}{z_H^3}}}+\frac{3 \delta  u^2 z_t^2 \left(1-\frac{u z_t}{z_H}\right)}{2 z_H^2 \left(1-\frac{u^3 z_t^3}{z_H^3}\right){}^{3/2}}\right) \, du -\frac{2LR^2}{z_{b}}.
\end{equation}
Once again using the identities given in \cref{identities}, the above integral reduces to following form
\begin{equation}
\begin{aligned}
 {\cal A}^{finite}
  &= \frac{LR^{2}}{z_t}\bigg[\sqrt{\pi}\frac{\Gamma(-\frac{1}{4})}{2\Gamma(\frac{1}{4})}+\sum_{n=1}^\infty \frac{\Gamma(n+\frac{1}{2})}{2\Gamma(n+1)}\frac{\Gamma(\frac{3n-1}{4})}{\Gamma(\frac{3n+1}{4})}\left(\frac{z_t}{z_H}\right)^{3n} \nonumber \\ 
  & +\frac{3\delta}{2}\sum_{n=0}^\infty \frac{\Gamma(n+\frac{3}{2})}{\Gamma(n+1)}\frac{\Gamma(\frac{3n+1}{4})}{\Gamma(\frac{3n+3}{4})}\left(\frac{z_t}{z_H}\right)^{3n+2} -\frac{3\delta}{2}\sum_{n=0}^\infty \frac{\Gamma(n+\frac{3}{2})}{\Gamma(n+1)}\frac{\Gamma(\frac{3n+2}{4})}{\Gamma(\frac{3n+4}{4})}\left(\frac{z_t}{z_H}\right)^{3n+3}\bigg].
  \end{aligned}
\end{equation}
After using the identity $\Gamma(n+1)=n\Gamma(n)$ in the above equation leads to the following
\begin{equation}\label{A-finite-high-4D-s}
\begin{aligned}
{\cal A}^{finite} &=\frac{LR^{2}}{z_t}\bigg[\sqrt{\pi}\frac{\Gamma(-\frac{1}{4})}{2\Gamma(\frac{1}{4})}+\sum_{n=1}^\infty\bigg(1+\frac{2}{3n-1}\bigg) \frac{\Gamma(n+\frac{1}{2})}{2\Gamma(n+1)}\frac{\Gamma(\frac{3n+3}{4})}{\Gamma(\frac{3n+5}{4})}\left(\frac{z_t}{z_H}\right)^{3n}\\
&+\frac{3\delta}{2}\sum_{n=0}^\infty\bigg(1+\frac{2}{3n+1}\bigg) \frac{\Gamma(n+\frac{3}{2})}{\Gamma(n+1)}\frac{\Gamma(\frac{3n+5}{4})}{\Gamma(\frac{3n+7}{4})}\left(\frac{z_t}{z_H}\right)^{3n+2} \\
 &-\frac{3\delta}{2}\sum_{n=0}^\infty\bigg(1+\frac{2}{3n+2}\bigg) \frac{\Gamma(n+\frac{3}{2})}{\Gamma(n+1)}\frac{\Gamma(\frac{3n+6}{4})}{\Gamma(\frac{3n+8}{4})}\left(\frac{z_t}{z_H}\right)^{3n+3}\bigg].
 \end{aligned}
\end{equation}
We can now use \cref{l-series} in above \cref{A-finite-high-4D-s} to simplify the extremal area as
\begin{equation}
\begin{aligned}
{\cal A}^{finite} &= \frac{LR^{2}}{z_t}\bigg[\sqrt{\pi}\frac{\Gamma(-\frac{1}{4})}{2\Gamma(\frac{1}{4})}+\frac{l}{z_t}+ \sum_{n=1}^\infty\bigg(\frac{1}{3n-1}\bigg) \frac{\Gamma(n+\frac{1}{2})}{\Gamma(n+1)}\frac{\Gamma(\frac{3n+3}{4})}{\Gamma(\frac{3n+5}{4})}\left(\frac{z_t}{z_H}\right)^{3n}\\
& +3\delta\sum_{n=0}^\infty\bigg(\frac{1}{3n+1}\bigg) \frac{\Gamma(n+\frac{3}{2})}{\Gamma(n+1)}\frac{\Gamma(\frac{3n+5}{4})}{\Gamma(\frac{3n+7}{4})}\left(\frac{z_t}{z_H}\right)^{3n+2} \\
&-3\delta\sum_{n=0}^\infty\bigg(\frac{1}{3n+2}\bigg) \frac{\Gamma(n+\frac{3}{2})}{\Gamma(n+1)}\frac{\Gamma(\frac{3n+6}{4})}{\Gamma(\frac{3n+8}{4})}\left(\frac{z_t}{z_H}\right)^{3n+3}\bigg]. 
 \end{aligned}
\end{equation}
The first series in the above equation goes as $ \frac{2}{3\sqrt{3}n^2} (z_t/z_H)^{3n}$ while the latter two series scale as  $\frac{2}{3\sqrt{3}n}(z_t/z_H)^{3n}$ for large $n$. Given that these series diverge at $\mathcal{O}(\epsilon)$ and beyond, we separate the divergent parts to obtain the following 
\begin{equation}
\begin{aligned}
{\cal A}^{finite} &= \frac{LR^{2}}{z_t}\Bigg[\sqrt{\pi}\frac{\Gamma(-\frac{1}{4})}{2\Gamma(\frac{1}{4})}+\frac{l}{z_t}+ \sum_{n=1}^\infty\bigg(\frac{1}{3n-1} \frac{\Gamma(n+\frac{1}{2})}{\Gamma(n+1)}\frac{\Gamma(\frac{3n+3}{4})}{\Gamma(\frac{3n+5}{4})}-\frac{2}{3\sqrt{3}n^2}\bigg)\left(\frac{z_t}{z_H}\right)^{3n}\\
&+3\delta\sum_{n=1}^\infty\bigg(\frac{1}{3n+1} \frac{\Gamma(n+\frac{3}{2})}{\Gamma(n+1)}\frac{\Gamma(\frac{3n+5}{4})}{\Gamma(\frac{3n+7}{4})}-\frac{2}{3\sqrt{3}n}\bigg)\left(\frac{z_t}{z_H}\right)^{3n+2} \\
&- 3\delta\sum_{n=1}^\infty\bigg(\frac{1}{3n+2} \frac{\Gamma(n+\frac{3}{2})}{\Gamma(n+1)}\frac{\Gamma(\frac{3n+6}{4})}{\Gamma(\frac{3n+8}{4})}-\frac{2}{3\sqrt{3}n}\bigg)\left(\frac{z_t}{z_H}\right)^{3n+3}\\
 &+ 3\delta\bigg(\frac{\Gamma(\frac{3}{2})\Gamma(\frac{5}{4})}{\Gamma(\frac{7}{4})} \left(\frac{z_t}{z_H}\right)^2 -\frac{(\Gamma(\frac{3}{2}))^2}{2}\left(\frac{z_t}{z_H}\right)^3\bigg) +\frac{2}{3\sqrt{3}}\mathrm{Li}_2\left[\left(\frac{z_t}{z_H}\right)^3\right]\\
 &-\frac{2\delta}{\sqrt{3}}(1-\frac{z_t}{z_H})\left(\frac{z_t}{z_H}\right)^2\log\left[1-\left(\frac{z_t}{z_H}\right)^3\right]\Bigg].
 \end{aligned}
\end{equation}
Upon substituting $z_t = z_H(1 - \epsilon)$ into the above expression and performing an expansion to order $\mathcal{O}(\epsilon)$ allow us to recast the result in the compact form as
\begin{equation}\label{A-finite-high-T}
{\cal A}^{ finite}=\frac{LR^{2}l}{z_H^2}+\frac{LR^{2}}{z_H}\Bigg[
A_0+\delta D_0
+\epsilon\Big(A_1+2\frac{l}{z_H}+\delta D_1
+\frac{2}{\sqrt{3}}(1-\delta)\ln\epsilon
\Big)
+\mathcal{O}(\epsilon^2)
\Bigg],
\end{equation}
where the constants $A_{0,1}, D_{0,1}$ are detailed in the appendix \ref{A-constant}.
Now using the RT formula, the HEE in the limit of high temperature and small backreaction can be expressed as follows
\begin{equation}\label{S-high-T}
S_A^{finite}=Ll \,s+\frac{LR^{2}}{4G_N^{4}z_H}\Bigg[
A_0+\delta D_0
+\epsilon\Big(A_1+2\frac{l}{z_H}+\delta D_1
+\frac{2}{\sqrt{3}}(1-\delta)\ln\epsilon
\Big)
+\mathcal{O}(\epsilon^2)
\Bigg].
\end{equation}
Here the term $s=\frac{R^{2}}{4G_N^4 z_H^2}$ corresponds to the entropy density of the black hole as can be seen from from the \cref{entropy-density}. In \cref{S-high-T}, the first term follows a volume law scaling (proportional to the subsystem area in the $(2+1)$ dimensional dual field theory) and grows with temperature indicating that it represents the thermal part of the entropy. The subleading terms scale with the length of the entangling surface separating subsystem $A$ from its complement and thus represents entanglement contributions at high temperatures. Notably, the $\epsilon$-corrections exhibit an exponential decay with temperature analogous to the Schwarzschild-AdS case although they are modified by a small $\delta$ term arising from the backreaction.

\section*{\textbf{Holographic Entanglement Negativity}}\label{sec:hen}
Entanglement negativity is a well established quantum information theoretic measure that captures the distillable quantum correlations in mixed states, making it particularly suitable for thermal or finite density systems where entanglement entropy alone fails to isolate genuine quantum correlations. Defined via the partial transpose of the reduced density matrix, negativity provides a computable upper bound on the distillable entanglement. Within the AdS/CFT framework, a holographic prescription for entanglement negativity in higher dimensions has been proposed, wherein suitable linear combinations of Ryu–Takayanagi surfaces encode the negativity of boundary mixed states \cite{Chaturvedi:2016rft,Jain:2017xsu,Jain:2018bai,KumarBasak:2020viv,Mondal:2021kzj,Afrasiar:2021hld}. This construction allows one to extend the study of mixed state entanglement to strongly coupled holographic plasmas, and in what follows we analyze its behavior in the deformed AdS black hole background to compare with the trends observed for HEE.

\section{\textbf{HEN for Adjacent Subsystems}}\label{sec:hen-adjacent}

Let us consider two adjacent rectangular strip subsystems $A_1$ and $A_2$ with respective lengths $l_1$ and $l_2$ described by
\begin{equation}
x\equiv x^1 \in\biggl[-\frac{l_j}{2},\frac{l_j}{2}\biggr],\quad x^i\in\biggl[-\frac{L}{2},\frac{L}{2}\biggr];\quad j=1,2,
\end{equation}
where $i=2,3,...,d-1$ and $L\to \infty$. The holographic prescription then requires an appropriate combination of the entanglement entropies of the individual regions and their union \cite{Jain:2017aqk,Jain:2017xsu}. This construction effectively isolates the quantum part of correlations by eliminating purely thermal contributions. In what follows, we will evaluate the holographic negativity for the adjacent strip configuration in the backreacted AdS black hole background and examine its dependence on quark density $\rho$, with particular focus on the low and high temperature limits, in order to contrast its behavior with that of the entanglement entropy discussed previously. Following the proposal given in \cite{Jain:2017xsu}, the HEN for the mixed state configuration of two adjacent subsystems is given by 
\begin{equation}
    \mathcal{E}=\frac{3}{16 G_N^{d+1}}(\mathcal{A}_1+\mathcal{A}_2-\mathcal{A}_{3}),
    \label{entn}
\end{equation}
Here, $\mathcal{A}_1$, $\mathcal{A}_2$ and $\mathcal{A}_3 = \mathcal{A}_1 + \mathcal{A}_2$ represent the areas of the corresponding RT surfaces, obtained using the prescription reviewed in \cref{sec:EErev}. To express the entanglement negativity in terms of boundary parameters, one needs to determine the turning point $z_t$ as a function of the strip width $l$ by solving \eqref{turning-point}. However, as discussed in \cref{sec:EErev}, this relation cannot be solved exactly for arbitrary $z_t$ and admits analytic solutions only in the low and high temperature regimes. Therefore, we evaluate the HEN in these two limiting cases using the corresponding results for the HEE.

\subsection{Low temperature limit}\label{sec:adjacent-low-discuss}
The low effective temperature regime corresponds to the condition $z_t \ll z_H$. In terms of the boundary parameter, this limit translates to $T_f l_j \ll 1$. By solving \cref{turning-point} perturbatively for $z_t$ in powers of $l$, one can obtain the corresponding expressions for the area functions $\mathcal{A}_1$, $\mathcal{A}_2$ and $\mathcal{A}_3$ in this regime, as outlined in \cite{Chakrabortty:2020ptb}. The resulting expressions for the area can be written from \cref{area-low-temp-ent} is given by

\begin{equation}
\label{lowS}
\begin{split}
\mathcal{A}_{j}=\frac{2}{d-2} \frac{L^{d-2} R^{d-1} }{\epsilon^{d-2}}&+R^{d-1}\left(\frac{L}{l_j}\right)^{d-2}S_0\Biggl[1+\rho S_2\left(\frac{4\pi T_{f}l_j}{d}\right)^{d-1}\\&+(1-\rho)S_1\left(\frac{4\pi T_{f}l_j}{d}\right)^d+\mathcal{O}\left((T_{f}l_j)^{2d-2}\right)\Biggr].
\end{split}
\end{equation}
Here, the index $j = 1,2,3$ labels the subsystems $A_1$, $A_2$ and $A_3$ corresponding to their respective strip widths. In \cref{lowS}, the first term with $\epsilon$ denotes the divergent contribution while $S_0$, $S_1$, and $S_2$ are constants depending only on the spacetime dimension $d$ and are given in \cref{Snotonetwo}.
Now using \eqref{entn} together with the low temperature expansion in \cref{lowS} for the area functionals, the HEN for two adjacent strip subsystems in the deformed AdS black hole background takes the following form
\begin{equation}
\begin{split}
  \mathcal{E} &= \frac{3}{8G_N^{d+1}(d-2)} \frac{L^{d-2} R^{d-1} }{\epsilon^{d-2}} + \frac{3R^{d-1}L^{d-2}}{16G_N^{d+1}}S_0\Biggl[ l_1^{2-d} + l_2^{2-d} - (l_1+l_2)^{2-d} \\
  &\qquad\qquad\qquad\qquad\qquad\qquad -\; 2 S_1(1-\rho)\, l_1 l_2\Big(\tfrac{4\pi T_f}{d}\Big)^{\!d} + \mathcal{O}\!\big((T_{f}l_j)^{2d-2}\big)\Biggr].
\end{split}
\label{eq:enlt-repeat}
\end{equation}
The following observations can be made from the above expression in \cref{eq:enlt-repeat} as follows:
\begin{itemize}
  \item  The first term is the standard ultraviolet divergence arising from short distance correlations near the entangling surface. The leading finite contribution (independent of $T_f$) is the familiar vacuum like short distance scaling for adjacent strip geometries and encodes the purely geometric contribution to negativity.

  \item The first nontrivial thermal/backreaction correction appears at order $T_f^{d}$ and is multiplied by the factor $(1-\rho)$. In low temperature regime $T_f^{\,d}$ term is the leading thermal correction and higher order terms scale as $(T_f l_j)^{2d-2}$ which are subleading for sufficiently small $T_f l_j$.

  \item The correction term proportional to $(1-\rho)T_f^{\,d}$ increases negativity. For fixed $T_f$ and subsystem sizes, the HEN therefore increases as the backreaction increases. Physically, this indicates that adding heavy flavor backreaction tends to enhance the correlations due to the addition of extra degrees of freedoms coming from the addition of backreaction.

\end{itemize}

The low temperature adjacent subsystems result in \cref{eq:enlt-repeat} demonstrates that in the near-vacuum regime, heavy-quark backreaction produces a leading thermal correction that raise the HEN. This is consistent with the qualitative expectation that additional matter content increases correlations and thereby enhance quantum entanglement as measured by negativity. We also observe that for $\rho = 0$, the expression in \cref{eq:enlt-repeat} simplifies to the case of an undeformed geometry and matches with the corresponding HEN for adjacent subsystems at low temperature limit \cite{Jain:2017xsu}.

\subsection{High temperature limit}
\label{sec:adjacent-high}

 In the regime of high effective temperature i.e when $T_f l_j \gg 1$, the HEN for the mixed state configuration of adjacent subsystems can be obtained using \cref{hiee} and \cref{entn} as follows
\begin{equation}
  \mathcal{E} = \; \frac{3\,L^{d-2} R^{d-1}}{8G_N^{d+1}}\left[\frac{1}{(d-2)\epsilon^{d-2}}+
  \biggl(\frac{4\pi T_f}{d}\biggr)^{\!d-2}\,
  \widetilde{S}(\rho,d)\right],
  \label{eq:enhtaj-repeat}
\end{equation}
where the dimensionless functions $\widetilde S$ is defined in \cref{stilde} and encode the backreaction dependence. In \cite{Chakrabortty:2020ptb} it has been shown that $\widetilde S(\rho,d)$ is a monotonically increasing function of $\rho$. From \cref{eq:enhtaj-repeat}, we observe the following:
\begin{itemize}

\item Unlike the HEE, the HEN is free from any volume dependent contribution. The terms proportional to the subsystem volume which encode classical and thermal correlations cancel exactly as expected. Since entanglement negativity quantifies purely quantum correlations in mixed states, it does not exhibit volume scaling and follow the area law behavior.
\item The quantity $\widetilde S$ is finite and encodes the entire nontrivial dependence on the backreaction parameter $\rho$ and the spacetime dimension $d$. As $\widetilde S$ increases monotonically with $\rho$, the holographic entanglement negativity correspondingly exhibits an enhancement with increasing backreaction.

\item Physically, this enhancement originates from the modification of the near horizon geometry due to matter backreaction. As the backreaction parameter $\rho$ increases, the geometric contribution entering the entanglement negativity acquires larger near horizon corrections which is reflected in the growth of $\widetilde S$.
    
\end{itemize}
In the high temperature regime, the HEN retains a single ultraviolet divergence matching the expected short distance behavior while remaining free from any volume law contribution due to the exact cancellation of volume dependent terms. The entire nontrivial dependence on the backreaction parameter $\rho$ and the spacetime dimension $d$ is encoded in the finite quantity $\widetilde S$ which increases monotonically with $\rho$. This enhancement reflects the growing influence of matter backreaction on the near horizon geometry leading to larger geometric contributions to the entanglement negativity. In the next section we will analyze the HEN for bipartite configuration.

\section{HEN for Bipartite System}
\label{sec:HEN-bipartite}

 Having analyzed the adjacent subsystems configuration, we now turn to the bipartite configuration, where the subsystem $A$ and its complement $A^c$ together form the entire boundary theory. In this setup, holographic entanglement negativity provides a sharp probe of mixed state quantum correlations allowing us to distinguish genuine quantum entanglement from classical and thermal correlations that dominate at high temperature.

To compute the HEN for bipartite system, we divide the $A^c$ in two parts denoted by $B_1$ and $B_2$ on the either side of $A$. We can define the subsystems as 
\begin{equation}
     B_1 \in [-\frac{l}{2},-L_1], \qquad A \in [-\frac{l}{2},\frac{l}{2}], \qquad B_2 \in [\frac{l}{2},L_2].
\end{equation}
The HEN for this mixed state configuration is then given by \cite{Chaturvedi:2016rft}
\begin{equation}
    \mathcal{E}=  \lim_{B\to A^c}\biggl[\frac{3}{16 G_N^{d+1}}(2\mathcal{A}_{A}+\mathcal{A}_{B_1}+\mathcal{A}_{B_2}- \mathcal{A}_{A\cup B_1}-\mathcal{A}_{A\cup B_2})\biggr],
\end{equation}
where $B=B_1\cup B_2$. We can choose $B_1=B_2$ which leads to the following
\begin{equation}\label{eq:ben}
    \mathcal{E}=\lim_{B\to A^c}\frac{3}{8 G_N^{d+1}}(\mathcal{A}_{A}+\mathcal{A}_{B_1}- \mathcal{A}_{A\cup B_1}),
\end{equation}
where $B\to A^c$ is the bipartite limit and can be imposed as $L_{1}\to \infty$.

\subsection{Low temperature analysis}\label{sec:HEN-bipartite-low}

In this section, we compute the HEN for a bipartite configuration in the low effective temperature limit defined by $T_fl\ll 1$. In the bulk, this regime implies that the turning point of the minimal surface for subsystem $A$ is located far from the black hole horizon. Conversely, for the subsystems $B_1$ and $A \cup B_1$ with lengths $(L_1 - l/2)$ and $(L_1 + l/2)$, the limit $B\to A^c$ $(L_1\to \infty)$ ensures that their respective minimal surfaces $\mathcal{A}_{B_1}$ and $\mathcal{A}_{A \cup B_1}$ probe the deep bulk and approach the horizon. By substituting the low temperature area $\mathcal{A}_{A}$ from \cref{HEE-lowT-final} and the high temperature limits for $\mathcal{A}_{B_1}$ and $\mathcal{A}_{A \cup B_1}$ from \cref{hiee} into \cref{eq:ben}, we obtain the HEN for bipartite configuration as follows
\begin{equation}
\begin{split}\label{E-bip-largeL}
\mathcal{E} =\frac{3R^{d-1}}{8G_N^{d+1}}\Bigg[\frac{2}{(d-2)}\left(\frac{L}{\epsilon}\right)^{d-2}
+S_0 \left(\frac{L}{l}\right)^{d-2}
+L^{d-2} \rho\,S_0 S_2 l \Big(\tfrac{4\pi T_f}{d}\Big)^{d-1} \\
+ L^{d-2}S_0 S_1(1-\rho)l^2\Big(\frac{4\pi T_f}{d}\Big)^{d} -V\Big(\tfrac{4\pi T_f}{d}\Big)^{d-1}
\Bigg],
\end{split}
\end{equation}
where $V=lL^{d-2}$ is the volume of the subsystem $A$ in a $d$ dimensional dual field theory. The above expression for the HEN in the low effective temperature regime can be written in the following form
\begin{equation}\label{en-substract}
    \mathcal{E}=\frac{3}{2}\left(S_A-S_A^{th}\right),
\end{equation}
where $S_A$ is the entanglement entropy for the subsystem $A$ given in \cref{HEE-lowT-final} and $S_A^{th}=\frac{R^{d-1}}{4G_N^{d+1}}\Big(\tfrac{4\pi T_f}{d}\Big)^{d-1}V$ is the thermal entropy associated with the subsystem $A$. The above relation in \cref{en-substract} demonstrates that at leading order, the HEN is obtained by subtracting the volume dependent thermal entropy from the HEE for the finite temperature mixed state of a single subsystem. This result aligns with quantum information theoretic expectations where entanglement negativity serves as an upper bound on distillable entanglement by isolating the purely quantum correlations in a mixed state.
This kind of relation has also been established in the literature at various scenarios \cite{Chaturvedi:2016rft, Mondal:2021kzj,Karan:2023hfk}.
One can draw the following conclusions from the above result:

\begin{itemize}
  \item The first term represents the UV divergent contribution similar to HEE. Within the remaining finite part, the dominant geometric scaling is governed by $l$ which recovers the characteristic short distance behavior of a strip in a pure AdS geometry.
  \item Unlike the adjacent case there are two $\rho$-dependent corrections with different parametric dependence first is  a $\rho$-linear correction that increses HEN with backreaction. This term enters with the same geometric scaling $l$ as the leading finite term. The second one is \emph{ $(1-\rho)$} and carries a factor of $l^{2}$ relative to the first correction and is parametrically higher order in $T_f$. This term reduces the negativity as backreaction increases.
  \item For the low temperature expansion and for sufficiently small strip width $l$, the $\rho$ linear term typically dominates over the $(1-\rho)$ term.
  \end{itemize}
 Overall, the backreaction effects enter through two distinct $\rho$-dependent corrections: a linear term that multiplicatively modifies the vacuum contribution and a subleading $(1-\rho)$ term that is parametrically suppressed for small strip widths and low temperatures. In this regime, the $\rho$-linear correction provides the dominant backreaction effect on the HEN.

\subsection{High temperature analysis}
\label{sec:HEN-bipartite-high}
In the limit of high effective temperature $(T_fl\gg 1)$, we combine the bipartite prescription \eqref{eq:ben} with the area expansions from \cref{hiee} to arrive at the following expression for the HEN as
\begin{equation}
\mathcal{E}=\frac{3\,L^{d-2} R^{d-1}}{4G_N^{d+1}}\left[\frac{1}{(d-2)\epsilon^{d-2}}+
  \biggl(\frac{4\pi T_f}{d}\biggr)^{\!d-2}\,
  \widetilde{S}(\rho,d)\right],
\label{E-bip-high-raw}
\end{equation}
where $\widetilde{S}(\rho,d)$ is defined in \cref{stilde} and it increases monotonically with the backreaction parameter $\rho$. The above expression can also be written in the following form
\begin{equation}
    \mathcal{E}=\frac{3}{2}\left(S_A-S_A^{th}\right).
\end{equation}
Consistent with the previous case, this behavior indicates the elimination of extensive thermal components from the negativity. The important observation in this regimes are as follows:

\begin{itemize}
  \item We note that the leading order expression for the HEN in \cref{E-bip-high-raw} depends only on the area of the entangling surface $L^{d-2}$. This area law scaling for the mixed state negativity is consistent with the findings reported in various systems like finite-temperature quantum spin models and two-dimensional harmonic lattices \cite{DeNobili:2016nmj,PhysRevE.93.022128}.
  \item The HEN increases with the backreaction parameter $\rho$ through the monotonic growth of the coefficient $\widetilde{S}(\rho,d)$. Apart from the ultraviolet divergent term, the leading finite contribution exhibits a temperature dependence scaling as $T_f^{\,d-2}$.
  \item Similar to the adjacent configuration, the enhancement of entanglement negativity with increasing backreaction reflects the presence of additional degrees of freedom in the system. However, unlike the HEE which scales with the subsystem volume and becomes insensitive to quantum correlations at high temperature, the entanglement negativity remains a reliable probe of quantum correlations in this regime.
  \end{itemize}
  Thus the bipartite analysis reinforces the distinctive role of HEN in isolating quantum correlations in mixed states. Even in regimes where thermal effects and backreaction are significant, the negativity continues to capture  entanglement features that remain sensitive to the underlying degrees of freedom thereby providing a complementary and more faithful diagnostic than other entanglement measures at high temperature.

\section{\textbf{HEN for two Disjoint Subsystems}}\label{sec:hen-disjoint}
We now analyze the HEN for the case of mixed state configuration of two disjoint subsystems. This configuration is described by two long rectangular strips $A_1$ and $A_2$ separated by an another strip $A_m$ as follows
\begin{equation}
x\equiv x^1 \in[-\frac{l_j}{2},\frac{l_j}{2}],\quad x^i \in [-\frac{L}{2},\frac{L}{2}];\quad i=2,3,\dots,(d-1),\quad j=1,2,m,
\end{equation}
where $L\gg l_1,l_2,l_m$ in order to neglect the corner effects. To determine the HEN for the two disjoint subsystems $A_1$ and $A_2$, we follow the the method outlined in \cite{KumarBasak:2020viv}. The HEN for the disjoint case can be expressed as
\begin{equation}
    \mathcal{E}=\frac{3}{16 G_N^{d+1}}(\mathcal{A}_{A_1\cup A_m}+\mathcal{A}_{A_m\cup A_2}-\mathcal{A}_{A_1\cup A_m\cup A_2}-\mathcal{A}_{A_{m}}),
    \label{enb}
\end{equation}
where $\mathcal{A}_{A_i\cup A_j}$ and $\mathcal{A}_{A_i\cup A_j\cup A_k}$ denotes the area of the RT surfaces for the subsystems $A_i\cup A_j$ and $A_i\cup A_j\cup A_k,$ respectively with $i,j,k=1,2,m$.
Now we will focus on two regimes corresponding to low and high effective temperature limit as done in previous section.
\subsection{Low temperature limit}
\label{sec:HEN-disjoint-low}
In the low effective temperature regime ($T_f l_j \ll 1$), we obtain the analytic expression for HEN of two disjoint subsystems by using \cref{HEE-lowT-final} in \cref{enb} as follows
\begin{equation}
\begin{split}
  \mathcal{E} = \frac{3R^{d-1}L^{d-2}}{16G_N^{d+1}} S_0 \Biggl[ & (l_1 + l_m)^{2-d} + (l_2 + l_m)^{2-d} - (l_1 + l_2 + l_m)^{2-d} - (l_m)^{2-d} \\
  & \hspace{0.3cm} -\, 2 S_1 (1-\rho)\, l_1 l_2 \left( \frac{4\pi T_f}{d} \right)^{\!d} + \mathcal{O}\!\left( (T_f l_j)^{2d-2} \right) \Biggr],
\end{split}
\label{enlt1}
\end{equation}
where $S_0$ and $S_1$ are $d$-dependent constants as defined earlier in \cref{Snotonetwo}. It is instructive to verify that in the adjacent limit $l_m \to 0$, the above expression reproduces the HEN for adjacent subsystems obtained in \cref{eq:enlt-repeat}. Further, the above result matches with the corresponding HEN for disjoint subsystems in the absence of the backreaction $(\rho\to 0)$ \cite{KumarBasak:2020viv}. From \cref{enlt1} one can observe that:
\begin{itemize}
\item In contrast to the adjacent and bipartite configurations, the ultraviolet divergent contributions cancel completely in this case, leaving the HEN finite. 
Consequently, the negativity scales with the area of the subsystem and obeys an area law.

  \item The leading contribution in \eqref{enlt1} is proportional to $S_0$ and depends solely on the geometric configuration of the disjoint intervals. This reflects the universal short-distance entanglement structure of the boundary theory.
  
  \item The backreaction enters explicitly through the correction term proportional to $(1-\rho)\,T_f^{\,d}$. Similar to the adjacent case, this term increases the negativity with increasing backreaction. This indicates that the presence of heavy quarks or extra degrees of freedoms enhances the quantum correlations at low temperature.
  
  \item The separation width $l_m$ plays a nontrivial role as $l_m$ increases, the negativity decreases due to reduced overlap of the entanglement wedges which is consistent with the expectation that large separations weaken correlations.
\end{itemize}
In summary, the low temperature behavior of HEN for disjoint subsystems captures both the universal short distance scaling and the effect of backreaction. In the next section, we will analyze the high temperature regime where the interplay between thermal contributions and $\rho$-dependent corrections leads to qualitatively different behavior.

\subsection{High temperature limit}
\label{sec:HEN-disjoint-high}
We now turn to the analysis of HEN for mixed state configuration of two disjoint subsystems in the high effective temperature regime under $l_m\ll l_{1,2}$ condition. In this case we utilize the high temperature result \cref{hiee} for all area functions appearing in \cref{enb} except $\mathcal{A}_m$, for which the low temperature result \cref{HEE-lowT-final} is applicable. Now  employing the general expressions from \cref{HEE-lowT-final,hiee} in \cref{enb} lead to the following expression for HEN as
\begin{equation}
\begin{aligned}
  \mathcal{E} = \frac{3 L^{d-2} R^{d-1}}{16G_N^{d+1}} \Biggl[ 2\biggl(\frac{4\pi T_f}{d}\biggr)^{d-2}\,
  \widetilde{S}(\rho,d)&+l_m\biggl(\frac{4\pi T_f}{d}\biggr)^{d-1}(1-S_0S_2\rho)
\\&- S_0S_1(1-\rho)\, l_m^{2}
\left(\frac{4\pi T_f}{d}\right)^d- S_0\,l_m^{2-d}\Biggl],
  \label{enht1}
  \end{aligned}
\end{equation}
The above expression is independent of $l_1$ and $l_2$ and contain only the width $l_m$ of the intermediate region $A_m$. The following are the main observations for this case: 
  \begin{itemize} 
  \item The term proportional to volume contributions from subsystems $A_1$ and $A_2$ get completely subtracted out as they are in large effective temperature regime.
  \item The first finite term scale as $T_f^{\,d-2}$ and encoding the nontrivial dependence on the backreaction and spacetime dimension through $\widetilde{S}(\rho,d)$. This term captures quantum correlations that persist beyond the leading thermal behavior and suggest an increase in correlation with backreaction.
  \item Next term is linear in the separation width $l_m$ and scales as $T_f^{\,d-1}$, indicating that at high temperature the entanglement negativity increases with separation between subsystems. The factor $(1-S_0S_2\rho)$ shows that increasing backreaction suppresses this contribution.
  
  \item The negative contributions proportional to $l_m^{2}$ and $l_m^{2-d}$ reflect the competing effects of thermal and geometric suppression at large separation. In particular, the $(1-\rho)\,l_m^{2}T_f^{\,d}$ term implies that increasing backreaction further increase the negativity.
\end{itemize}
Overall, the high-temperature entanglement negativity for disjoint subsystems exhibits a rich interplay between thermal effects, geometric separation and matter backreaction. While contributions proportional to $\widetilde{S}(\rho,d)$ encode persistent quantum correlations that are enhanced by backreaction, the separation-dependent terms reveal competing mechanisms that either decrease or amplify the negativity depending on the parametric regime. In the following section, we analyze how these competing contributions determine a critical separation beyond which the entanglement negativity of the disjoint configuration vanishes.

 \section{Critical Separation and Entanglement Disentanglement Transition}\label{sec:critical-separation}
In this section we will compute the critical separation between two subregions in disjoint configuration. In \cite{Chakrabortty:2020ptb}, the effect of backreaction on critical separation between two subsystems is studied using holographic mutual information and it was observed that the critical separation increases with backreaction. Since the HEN is also an entanglement measure for quantum correlations, we can use it to detect the critical separation or phase transition from entangled to disentangled phase. The critical separation is defined as the value of the intermediate width at which the HEN vanishes. The HEN for disjoint subsystems from \cref{enb} can be rewritten as
\begin{equation}
\mathcal{E}
=\frac{3}{4}\Big(
S(l_1+l_m)+S(l_2+l_m)-S(l_1+l_2+l_m)-S(l_m)
\Big).
\label{eq:neg-disjoint}
\end{equation}
At critical width $l_m=l_c$, we have
\begin{equation}
\mathcal{E}(l_c)=0.
\label{eq:lc-def}
\end{equation}
We focus on the physically relevant regime $l_c \ll l_1,l_2$, where the entropies associated with the larger subsystems probe the high temperature limit while the entropy corresponding to the separating region remains in the low temperature regime. Accordingly,
\begin{equation}
\begin{split}
& S(l_1+l_c)=S^{\text{high}}(l_1+l_c), \qquad
S(l_2+l_c)=S^{\text{high}}(l_2+l_c),\\
& S(l_1+l_2+l_c)=S^{\text{high}}(l_1+l_2+l_c), \qquad
S(l_c)=S^{\text{low}}(l_c).
\end{split}
\label{eq:hybrid}
\end{equation}
From the high and low temperature expansions of the HEE and imposing \cref{eq:lc-def} we obtain, 
\begin{equation}
\begin{split}
2 \widetilde{S}(\rho,d) \left(\frac{4\pi}{d}\right)^{d-2} (T_f l_c)^{d-2}
+
(1-S_0 S_2 \rho)& \left(\frac{4\pi}{d}\right)^{d-1} (T_f l_c)^{d-1}\\&+S_0S_1(1-\rho)\left(\frac{4\pi}{d}\right)^{d-2}(T_f l_c)^{d}
=
S_0 .
\end{split}
\label{eq:lc_general}
\end{equation} 
We now focus on the regime $T_f l_c \ll 1$
which corresponds to small separations compared to the thermal scale. In this limit, the $(T_f l_c)^{d-2}$ term dominates over the $(T_f l_c)^{d-1}$ and $(T_f l_c)^{d}$ terms, and \cref{eq:lc_general} reduces at leading order to
\begin{equation}
2 \widetilde{S}(\rho,d) \left(\frac{4\pi}{d}\right)^{d-2} (T_f l_c)^{d-2}
\simeq
S_0 .
\end{equation}
Solving for the critical separation, we obtain
\begin{equation}
l^{(0)}_c
\simeq
\frac{1}{T_f}
\left[
\frac{S_0}{2\widetilde{S}(\rho,d)}
\left(\frac{d}{4\pi}\right)^{d-2}
\right]^{\frac{1}{d-2}} .
\label{eq:lc_solution}
\end{equation}
Since $\widetilde{S}(\rho,d)$ is a monotonically increasing function of the backreaction parameter $\rho$ and $S_0$ is negative constant, \cref{eq:lc_solution} as noted in \cite{Chakrabortty:2020ptb}, $\widetilde{S}(0,4)=-0.33$ and $\widetilde{S}(1,4)=-0.024$ therefore it immediately implies that in the range of backreaction $\rho=0$ to $\rho=1$ we have
\begin{equation}
\frac{\partial l_c}{\partial \rho} > 0 .
\end{equation}
However, to see this increasing behavior for full range of $\rho$ we need to include all terms. Thus, in the regime $T_f l_c \ll 1$, the critical separation extracted from entanglement negativity increases with increasing backreaction up to a certain value of $\rho$ after a particular value of backreaction critical separation becomes imaginary which requires further computation with subleading corrections in \cref{eq:lc_general}.

This behavior is similar with the critical separation obtained from mutual information in \cite{Chakrabortty:2020ptb}.
The critical separation characterizes the spatial range over which distillable quantum correlations persist. Matter backreaction enhances thermal mixing and accelerates the decay of quantum correlations with distance, causing the negativity to vanish. While mutual information diagnoses the persistence of total correlations, entanglement negativity provides a sharper probe of the decay of quantum entanglement in strongly coupled holographic systems with matter backreaction.

\section{Summary and Discussion}\label{sec:summary}

To summarize, we have investigated the entanglement thermodynamics and holographic entanglement negativity (HEN) in a strongly coupled large $N_c$ gauge theory at finite temperature backreacted by the heavy static fundamental quarks. This field theory is holographically dual to a deformed AdS black hole geometry where the deformation/backreaction is sourced by the large number of long static strings (string cloud). We began by revisiting the holographic entanglement entropy in both the low and high temperature regimes. The analysis of HEE in low temperature limit allowed us to identify a first law like relation for the entanglement entropy. Furthermore, we provided a review of the entanglement wedge cross section and discussed its implications for reflected entropy.

Subsequently, we explored the behavior of HEN for adjacent, bipartite and disjoint subsystem configurations. The main focus of this work is to understand how the backreaction modifies quantum correlations across different thermal regimes and how these results compare with the corresponding behavior of HEE. Our analysis reveals a striking regime dependence: for adjacent subsystems case, the low temperature expansion produces a correction proportional to $(1-\rho)T_f^{\,d}$ which enhances HEN and thus confirms that backreaction increases the distillable entanglement in the near vacuum phase. In the high temperature limit, the relevant $\widetilde S(\rho,d)$ coefficients increase with $\rho$ leading to an enhancement of HEN. For the bipartite configuration, the HEN shows increasing behavior with backreaction in low and high temperature limits, and isolate the thermal entropy.
For disjoint subsystems, a similar increasing trend emerges in the low $T_f$ regime is dominated by $(1-\rho)$ terms, while in the high $T_f$ due to monotonically increasing $\widetilde S(\rho,d)$ coefficients. In contrast to the holographic entanglement entropy and mutual information which capture total correlations and become insensitive to quantum entanglement at high temperature, the more reliable measure provided by holographic entanglement negativity successfully captures the enhancement of distillable quantum correlations induced by backreaction.
 From our analysis of the critical separation for the disjoint configuration using holographic entanglement negativity, we find that the critical separation increases with increasing backreaction. This behavior is consistent with the enhancement of quantum correlations induced by backreaction, implying that a larger spatial separation is required to completely destroy entanglement. In this sense, stronger backreaction shifts the transition between the entangled and disentangled phases to larger separations, highlighting the role of matter backreaction in shaping the entanglement structure of strongly coupled holographic plasmas.

Overall, the observations made here provide a deeper understanding of entanglement and
related phenomena in strongly coupled conformal field theories with backreaction at finite temperature. These results establish that the influence of adding the heavy flavor backreaction is universal and increases the entanglement in the theory through the extra degrees of freedom. This highlights HEN as a sharper diagnostic of the balance between quantum and thermal correlations than HEE alone. Moreover, our findings are consistent with the previous studies on the effect of backreaction on entanglement. Future directions include performing a systematic numerical study in the intermediate regime $T_f l\sim \mathcal{O}(1)$ to map out crossover behavior, exploring time-dependent quenches to connect with thermalization dynamics and extending the analysis to higher-derivative gravity or string-theoretic embeddings to quantify $1/N$ and $\alpha'$ corrections. Such studies could shed further light on the microscopic structure of entanglement in strongly coupled plasmas with flavor backreaction and on the universal features of entanglement negativity in holographic duals.

\section*{Acknowledgments}
S.P would like to thank Prof. Hemwati Nandan for his support during the visit to HNB Garhwal University. H.P acknowledges the support of this work by NCTS.

\begin{appendices}
  \section{Numerical coefficients}\label{A-constant}
The constants used in \cref{A-finite-high-T} are given by
\begin{equation}
\begin{aligned}
A_0 &= 
\sqrt{\pi}\,\frac{\Gamma\!\left(-\tfrac14\right)}
     {2\,\Gamma\!\left(\tfrac14\right)}+ W_{a,0}+\frac{\pi^2}{9\sqrt{3}},\quad A_1 =A_0 -\,W_{a,1}-\frac{2}{\sqrt{3}}+\frac{2\ln 3}{\sqrt{3}}
\\
D_0 &= 3\,W_{b,0}-3\,W_{c,0}+3\big(G_1 - G_2\big), \quad G_1=\frac{\Gamma(\tfrac32)\Gamma(\tfrac54)}{\Gamma(\tfrac74)},
\quad 
G_2=\frac{(\Gamma(\tfrac32))^2}{2}\\
D_1 &= D_0-3\,W_{b,1}+3\,W_{c,1}+3(-2G_1+3G_2)-\frac{2\ln 3}{\sqrt{3}}\\
W_{a,0} &= \sum_{n=1}^{\infty} \tilde a_n, \quad W_{a,1} = \sum_{n=1}^{\infty} (3n)\,\tilde a_n,\quad
W_{b,0} = \sum_{n=1}^{\infty} \tilde b_n,  \quad W_{b,1} = \sum_{n=1}^{\infty} (3n+2)\,\tilde b_n,
\\
W_{c,0}&= \sum_{n=1}^{\infty} \tilde c_n, \quad W_{c,1} = \sum_{n=1}^{\infty} (3n+3)\,\tilde c_n\\
\tilde a_n &= 
\frac{1}{3n-1}
\frac{\Gamma(n+\tfrac12)}{\Gamma(n+1)}
\frac{\Gamma\!\left(\tfrac{3n+3}{4}\right)}
     {\Gamma\!\left(\tfrac{3n+5}{4}\right)} - \frac{2}{3\sqrt{3}\,n^2},\quad
\tilde b_n = 
\frac{1}{3n+1}
\frac{\Gamma(n+\tfrac32)}{\Gamma(n+1)}
\frac{\Gamma\!\left(\tfrac{3n+5}{4}\right)}
     {\Gamma\!\left(\tfrac{3n+7}{4}\right)} - \frac{2}{3\sqrt{3}\,n},\\
\tilde c_n &= 
\frac{1}{3n+2}
\frac{\Gamma(n+\tfrac32)}{\Gamma(n+1)}
\frac{\Gamma\!\left(\tfrac{3n+6}{4}\right)}
     {\Gamma\!\left(\tfrac{3n+8}{4}\right)} - \frac{2}{3\sqrt{3}\,n}.
\end{aligned}
\end{equation}

\end{appendices}

\bibliographystyle{JHEP}
\bibliography{NegQCM}

@article{Dong:2024gud,
    author = "Dong, Xi and Kudler-Flam, Jonah and Rath, Pratik",
    title = "{Entanglement negativity and replica symmetry breaking in general holographic states}",
    eprint = "2409.13009",
    archivePrefix = "arXiv",
    primaryClass = "hep-th",
    doi = "10.1007/JHEP01(2025)022",
    journal = "JHEP",
    volume = "01",
    pages = "022",
    year = "2025"
}

@article{Dong:2021clv,
    author = "Dong, Xi and Qi, Xiao-Liang and Walter, Michael",
    title = "{Holographic entanglement negativity and replica symmetry breaking}",
    eprint = "2101.11029",
    archivePrefix = "arXiv",
    primaryClass = "hep-th",
    doi = "10.1007/JHEP06(2021)024",
    journal = "JHEP",
    volume = "06",
    pages = "024",
    year = "2021"
}

@article{KumarBasak:2020ams,
    author = "Kumar Basak, Jaydeep and Basu, Debarshi and Malvimat, Vinay and Parihar, Himanshu and Sengupta, Gautam",
    title = "{Islands for entanglement negativity}",
    eprint = "2012.03983",
    archivePrefix = "arXiv",
    primaryClass = "hep-th",
    doi = "10.21468/SciPostPhys.12.1.003",
    journal = "SciPost Phys.",
    volume = "12",
    number = "1",
    pages = "003",
    year = "2022"
}

@article{Lunin:2025yth,
    author = "Lunin, Oleg",
    title = "{Gravitational Waves in the Myers-Perry Geometry}",
    eprint = "2510.14417",
    archivePrefix = "arXiv",
    primaryClass = "hep-th",
    month = "10",
    year = "2025"
}

@article{PhysRevE.93.022128,
  title = {Nonzero-temperature entanglement negativity of quantum spin models: Area law, linked cluster expansions, and sudden death},
  author = {Sherman, Nicholas E. and Devakul, Trithep and Hastings, Matthew B. and Singh, Rajiv R. P.},
  journal = {Phys. Rev. E},
  volume = {93},
  issue = {2},
  pages = {022128},
  numpages = {10},
  year = {2016},
  month = {Feb},
  publisher = {American Physical Society},
  doi = {10.1103/PhysRevE.93.022128},
  url = {https://link.aps.org/doi/10.1103/PhysRevE.93.022128}
}

@article{DeNobili:2016nmj,
    author = "De Nobili, Cristiano and Coser, Andrea and Tonni, Erik",
    title = "{Entanglement negativity in a two dimensional harmonic lattice: Area law and corner contributions}",
    eprint = "1604.02609",
    archivePrefix = "arXiv",
    primaryClass = "cond-mat.stat-mech",
    doi = "10.1088/1742-5468/2016/08/083102",
    journal = "J. Stat. Mech.",
    volume = "1608",
    number = "8",
    pages = "083102",
    year = "2016"
}

@article{Jeong:2019xdr,
    author = "Jeong, Hyun-Sik and Kim, Keun-Young and Nishida, Mitsuhiro",
    title = "{Reflected Entropy and Entanglement Wedge Cross Section with the First Order Correction}",
    eprint = "1909.02806",
    archivePrefix = "arXiv",
    primaryClass = "hep-th",
    doi = "10.1007/JHEP12(2019)170",
    journal = "JHEP",
    volume = "12",
    pages = "170",
    year = "2019"
}

@article{Jokela:2019ebz,
    author = {Jokela, Niko and P{\"o}nni, Arttu},
    title = "{Notes on entanglement wedge cross sections}",
    eprint = "1904.09582",
    archivePrefix = "arXiv",
    primaryClass = "hep-th",
    reportNumber = "HIP-2019-10/TH",
    doi = "10.1007/JHEP07(2019)087",
    journal = "JHEP",
    volume = "07",
    pages = "087",
    year = "2019"
}

@article{Umemoto:2018jpc,
    author = "Umemoto, Koji and Zhou, Yang",
    title = "{Entanglement of Purification for Multipartite States and its Holographic Dual}",
    eprint = "1805.02625",
    archivePrefix = "arXiv",
    primaryClass = "hep-th",
    reportNumber = "YITP-18-41",
    doi = "10.1007/JHEP10(2018)152",
    journal = "JHEP",
    volume = "10",
    pages = "152",
    year = "2018"
}

@article{Takayanagi:2017knl,
    author = "Takayanagi, Tadashi and Umemoto, Koji",
    title = "{Entanglement of purification through holographic duality}",
    eprint = "1708.09393",
    archivePrefix = "arXiv",
    primaryClass = "hep-th",
    reportNumber = "YITP-17-89, IPMU17-0115",
    doi = "10.1038/s41567-018-0075-2",
    journal = "Nature Phys.",
    volume = "14",
    number = "6",
    pages = "573--577",
    year = "2018"
}

@article{Jiang:2024ijx,
    author = "Jiang, Xin and Wang, Peng and Wu, Houwen and Yang, Haitang",
    title = "{Alternative to purification in conformal field theory}",
    eprint = "2406.09033",
    archivePrefix = "arXiv",
    primaryClass = "hep-th",
    reportNumber = "CTU-SCU/2024005",
    doi = "10.1103/PhysRevD.111.L021902",
    journal = "Phys. Rev. D",
    volume = "111",
    number = "2",
    pages = "L021902",
    year = "2025"
}

@article{BabaeiVelni:2019pkw,
    author = "Babaei Velni, Komeil and Mohammadi Mozaffar, M. Reza and Vahidinia, M. H.",
    title = "{Some Aspects of Entanglement Wedge Cross-Section}",
    eprint = "1903.08490",
    archivePrefix = "arXiv",
    primaryClass = "hep-th",
    reportNumber = "IPM/P-2019/007",
    doi = "10.1007/JHEP05(2019)200",
    journal = "JHEP",
    volume = "05",
    pages = "200",
    year = "2019"
}

@article{Tamaoka:2018ned,
    author = "Tamaoka, Kotaro",
    title = "{Entanglement Wedge Cross Section from the Dual Density Matrix}",
    eprint = "1809.09109",
    archivePrefix = "arXiv",
    primaryClass = "hep-th",
    reportNumber = "OU-HET-979",
    doi = "10.1103/PhysRevLett.122.141601",
    journal = "Phys. Rev. Lett.",
    volume = "122",
    number = "14",
    pages = "141601",
    year = "2019"
}

@article{Jain:2020rbb,
    author = "Jain, Parul and Mahapatra, Subhash",
    title = "{Mixed state entanglement measures as probe for confinement}",
    eprint = "2010.07702",
    archivePrefix = "arXiv",
    primaryClass = "hep-th",
    doi = "10.1103/PhysRevD.102.126022",
    journal = "Phys. Rev. D",
    volume = "102",
    pages = "126022",
    year = "2020"
}

@article{Jain:2022hxl,
    author = "Jain, Parul and Jena, Siddhi Swarupa and Mahapatra, Subhash",
    title = "{Holographic confining-deconfining gauge theories and entanglement measures with a magnetic field}",
    eprint = "2209.15355",
    archivePrefix = "arXiv",
    primaryClass = "hep-th",
    reportNumber = "APCTP Pre2022 - 023",
    doi = "10.1103/PhysRevD.107.086016",
    journal = "Phys. Rev. D",
    volume = "107",
    number = "8",
    pages = "086016",
    year = "2023"
}

@article{Jain:2022csf,
    author = "Jain, Parul and Jokela, Niko and Jarvinen, Matti and Mahapatra, Subhash",
    title = "{Bounding entanglement wedge cross sections}",
    eprint = "2211.07671",
    archivePrefix = "arXiv",
    primaryClass = "hep-th",
    reportNumber = "APCTP Pre2022 - 24 , HIP-2022-27/TH",
    doi = "10.1007/JHEP03(2023)102",
    journal = "JHEP",
    volume = "03",
    pages = "102",
    year = "2023"
}

@article{Basu:2022nds,
    author = "Basu, Debarshi and Parihar, Himanshu and Raj, Vinayak and Sengupta, Gautam",
    title = "{Entanglement negativity, reflected entropy, and anomalous gravitation}",
    eprint = "2202.00683",
    archivePrefix = "arXiv",
    primaryClass = "hep-th",
    doi = "10.1103/PhysRevD.105.086013",
    journal = "Phys. Rev. D",
    volume = "105",
    number = "8",
    pages = "086013",
    year = "2022",
    note = "[Erratum: Phys.Rev.D 105, 129902 (2022)]"
}

@article{Kudler-Flam:2018qjo,
    author = "Kudler-Flam, Jonah and Ryu, Shinsei",
    title = "{Entanglement negativity and minimal entanglement wedge cross sections in holographic theories}",
    eprint = "1808.00446",
    archivePrefix = "arXiv",
    primaryClass = "hep-th",
    doi = "10.1103/PhysRevD.99.106014",
    journal = "Phys. Rev. D",
    volume = "99",
    number = "10",
    pages = "106014",
    year = "2019"
}

@article{Kusuki:2019zsp,
    author = "Kusuki, Yuya and Kudler-Flam, Jonah and Ryu, Shinsei",
    title = "{Derivation of holographic negativity in AdS$_3$/CFT$_2$}",
    eprint = "1907.07824",
    archivePrefix = "arXiv",
    primaryClass = "hep-th",
    reportNumber = "YITP-19-65",
    doi = "10.1103/PhysRevLett.123.131603",
    journal = "Phys. Rev. Lett.",
    volume = "123",
    number = "13",
    pages = "131603",
    year = "2019"
}

@article{KumarBasak:2020eia,
    author = "Kumar Basak, Jaydeep and Malvimat, Vinay and Parihar, Himanshu and Paul, Boudhayan and Sengupta, Gautam",
    title = "{On Minimal Entanglement Wedge Cross Section for Holographic Entanglement Negativity}",
    eprint = "2002.10272",
    archivePrefix = "arXiv",
    primaryClass = "hep-th",
    doi = "10.3390/universe10030125",
    journal = "Universe",
    volume = "10",
    number = "3",
    pages = "125",
    year = "2024"
}

@article{KumarBasak:2021lwm,
    author = "Kumar Basak, Jaydeep and Parihar, Himanshu and Paul, Boudhayan and Sengupta, Gautam",
    title = "{Covariant holographic negativity from the entanglement wedge in AdS3/CFT2}",
    eprint = "2102.05676",
    archivePrefix = "arXiv",
    primaryClass = "hep-th",
    doi = "10.1016/j.physletb.2022.137451",
    journal = "Phys. Lett. B",
    volume = "834",
    pages = "137451",
    year = "2022"
}

@article{Chaturvedi:2016rcn,
    author = "Chaturvedi, Pankaj and Malvimat, Vinay and Sengupta, Gautam",
    title = "{Holographic Quantum Entanglement Negativity}",
    eprint = "1609.06609",
    archivePrefix = "arXiv",
    primaryClass = "hep-th",
    doi = "10.1007/JHEP05(2018)172",
    journal = "JHEP",
    volume = "05",
    pages = "172",
    year = "2018"
}

@article{Chaturvedi:2016rft,
    author = "Chaturvedi, Pankaj and Malvimat, Vinay and Sengupta, Gautam",
    title = "{Entanglement negativity, Holography and Black holes}",
    eprint = "1602.01147",
    archivePrefix = "arXiv",
    primaryClass = "hep-th",
    doi = "10.1140/epjc/s10052-018-5969-8",
    journal = "Eur. Phys. J. C",
    volume = "78",
    number = "6",
    pages = "499",
    year = "2018"
}

@article{Jain:2017xsu,
    author = "Jain, Parul and Malvimat, Vinay and Mondal, Sayid and Sengupta, Gautam",
    title = "{Holographic entanglement negativity for adjacent subsystems in AdS$_{d+1}$/CFT$_{d}$}",
    eprint = "1708.00612",
    archivePrefix = "arXiv",
    primaryClass = "hep-th",
    doi = "10.1140/epjp/i2018-12113-0",
    journal = "Eur. Phys. J. Plus",
    volume = "133",
    number = "8",
    pages = "300",
    year = "2018"
}

@article{Jain:2018bai,
    author = "Jain, Parul and Malvimat, Vinay and Mondal, Sayid and Sengupta, Gautam",
    title = "{Holographic Entanglement Negativity for Conformal Field Theories with a Conserved Charge}",
    eprint = "1804.09078",
    archivePrefix = "arXiv",
    primaryClass = "hep-th",
    doi = "10.1140/epjc/s10052-018-6383-y",
    journal = "Eur. Phys. J. C",
    volume = "78",
    number = "11",
    pages = "908",
    year = "2018"
}

@article{KumarBasak:2020viv,
    author = "Kumar Basak, Jaydeep and Parihar, Himanshu and Paul, Boudhayan and Sengupta, Gautam",
    title = "{Holographic entanglement negativity for disjoint subsystems in AdSd+1/CFTd}",
    eprint = "2001.10534",
    archivePrefix = "arXiv",
    primaryClass = "hep-th",
    doi = "10.1142/S0217751X24501409",
    journal = "Int. J. Mod. Phys. A",
    volume = "39",
    number = "32",
    pages = "2450140",
    year = "2024"
}

@article{Mondal:2021kzj,
    author = "Mondal, Sayid and Paul, Boudhayan and Sengupta, Gautam and Sharma, Punit",
    title = "{Holographic entanglement negativity for a single subsystem in conformal field theories with a conserved charge}",
    eprint = "2102.05848",
    archivePrefix = "arXiv",
    primaryClass = "hep-th",
    reportNumber = "CYCU-HEP-21-01",
    doi = "10.1088/1751-8121/acfb52",
    journal = "J. Phys. A",
    volume = "56",
    number = "42",
    pages = "425402",
    year = "2023"
}

@article{Afrasiar:2021hld,
    author = "Afrasiar, Mir and Kumar Basak, Jaydeep and Raj, Vinayak and Sengupta, Gautam",
    title = "{Holographic Entanglement Negativity for Disjoint Subsystems in Conformal Field Theories with a Conserved Charge}",
    eprint = "2106.14918",
    archivePrefix = "arXiv",
    primaryClass = "hep-th",
    month = "6",
    year = "2021"
}

@article{Jain:2017aqk,
    author = "Jain, Parul and Malvimat, Vinay and Mondal, Sayid and Sengupta, Gautam",
    title = "{Holographic entanglement negativity conjecture for adjacent intervals in $AdS_3/CFT_2$}",
    eprint = "1707.08293",
    archivePrefix = "arXiv",
    primaryClass = "hep-th",
    doi = "10.1016/j.physletb.2019.04.037",
    journal = "Phys. Lett. B",
    volume = "793",
    pages = "104--109",
    year = "2019"
}

@article{Chaturvedi:2017znc,
    author = "Chaturvedi, Pankaj and Malvimat, Vinay and Sengupta, Gautam",
    title = "{Covariant holographic entanglement negativity}",
    eprint = "1611.00593",
    archivePrefix = "arXiv",
    primaryClass = "hep-th",
    doi = "10.1140/epjc/s10052-018-6259-1",
    journal = "Eur. Phys. J. C",
    volume = "78",
    number = "9",
    pages = "776",
    year = "2018"
}

@article{Jain:2017uhe,
    author = "Jain, Parul and Malvimat, Vinay and Mondal, Sayid and Sengupta, Gautam",
    title = "{Covariant holographic entanglement negativity for adjacent subsystems in AdS$_3$ /CFT$_2$}",
    eprint = "1710.06138",
    archivePrefix = "arXiv",
    primaryClass = "hep-th",
    doi = "10.1016/j.nuclphysb.2019.114683",
    journal = "Nucl. Phys. B",
    volume = "945",
    pages = "114683",
    year = "2019"
}

@article{Malvimat:2018txq,
    author = "Malvimat, Vinay and Mondal, Sayid and Paul, Boudhayan and Sengupta, Gautam",
    title = "{Holographic entanglement negativity for disjoint intervals in $AdS_3/CFT_2$}",
    eprint = "1810.08015",
    archivePrefix = "arXiv",
    primaryClass = "hep-th",
    doi = "10.1140/epjc/s10052-019-6693-8",
    journal = "Eur. Phys. J. C",
    volume = "79",
    number = "3",
    pages = "191",
    year = "2019"
}

@article{Malvimat:2018ood,
    author = "Malvimat, Vinay and Mondal, Sayid and Paul, Boudhayan and Sengupta, Gautam",
    title = "{Covariant holographic entanglement negativity for disjoint intervals in $AdS_3/CFT_2$}",
    eprint = "1812.03117",
    archivePrefix = "arXiv",
    primaryClass = "hep-th",
    doi = "10.1140/epjc/s10052-019-7032-9",
    journal = "Eur. Phys. J. C",
    volume = "79",
    number = "6",
    pages = "514",
    year = "2019"
}

@article{Balasubramanian:1999re,
    author = "Balasubramanian, Vijay and Kraus, Per",
    title = "{A Stress tensor for Anti-de Sitter gravity}",
    eprint = "hep-th/9902121",
    archivePrefix = "arXiv",
    reportNumber = "HUTP-99-A002, EFI-99-6, NSF-ITP-98-132",
    doi = "10.1007/s002200050764",
    journal = "Commun. Math. Phys.",
    volume = "208",
    pages = "413--428",
    year = "1999"
}

@article{Mishra:2015cpa,
    author = "Mishra, Rohit and Singh, Harvendra",
    title = "{Perturbative entanglement thermodynamics for AdS spacetime: Renormalization}",
    eprint = "1507.03836",
    archivePrefix = "arXiv",
    primaryClass = "hep-th",
    doi = "10.1007/JHEP10(2015)129",
    journal = "JHEP",
    volume = "10",
    pages = "129",
    year = "2015"
}

@article{Sun:2016til,
    author = "Sun, Yuan and Xu, Hao and Zhao, Liu",
    title = "{Thermodynamics and holographic entanglement entropy for spherical black holes in 5D Gauss-Bonnet gravity}",
    eprint = "1606.06531",
    archivePrefix = "arXiv",
    primaryClass = "gr-qc",
    doi = "10.1007/JHEP09(2016)060",
    journal = "JHEP",
    volume = "09",
    pages = "060",
    year = "2016"
}

@article{OBannon:2016exv,
    author = "O'Bannon, Andy and Probst, Jonas and Rodgers, Ronnie and Uhlemann, Christoph F.",
    title = "{First law of entanglement rates from holography}",
    eprint = "1612.07769",
    archivePrefix = "arXiv",
    primaryClass = "hep-th",
    doi = "10.1103/PhysRevD.96.066028",
    journal = "Phys. Rev. D",
    volume = "96",
    number = "6",
    pages = "066028",
    year = "2017"
}

@article{McCarthy:2017amh,
    author = "McCarthy, Fiona and Kubiz{\v{n}}{\'a}k, David and Mann, Robert B.",
    title = "{Breakdown of the equal area law for holographic entanglement entropy}",
    eprint = "1708.07982",
    archivePrefix = "arXiv",
    primaryClass = "hep-th",
    doi = "10.1007/JHEP11(2017)165",
    journal = "JHEP",
    volume = "11",
    pages = "165",
    year = "2017"
}

@article{Saha:2019ado,
    author = "Saha, Ashis and Gangopadhyay, Sunandan and Saha, Jyoti Prasad",
    title = "{Holographic entanglement entropy and generalized entanglement temperature}",
    eprint = "1906.03159",
    archivePrefix = "arXiv",
    primaryClass = "hep-th",
    doi = "10.1103/PhysRevD.100.106008",
    journal = "Phys. Rev. D",
    volume = "100",
    number = "10",
    pages = "106008",
    year = "2019"
}

@article{Maulik:2020tzm,
    author = "Maulik, Sabyasachi and Singh, Harvendra",
    title = "{Entanglement entropy and the first law at third order for boosted black branes}",
    eprint = "2012.09530",
    archivePrefix = "arXiv",
    primaryClass = "hep-th",
    doi = "10.1007/JHEP04(2021)065",
    journal = "JHEP",
    volume = "04",
    pages = "065",
    year = "2021"
}

@article{Nadi:2019bqu,
    author = "Nadi, Hamideh and Mirza, Behrouz and Sherkatghanad, Zeinab and Mirzaiyan, Zahra",
    title = "{Holographic entanglement first law for $d$ + 1 dimensional rotating cylindrical black holes}",
    eprint = "1904.11344",
    archivePrefix = "arXiv",
    primaryClass = "gr-qc",
    doi = "10.1016/j.nuclphysb.2019.114822",
    journal = "Nucl. Phys. B",
    volume = "949",
    pages = "114822",
    year = "2019"
}

@article{Blanco:2013joa,
    author = "Blanco, David D. and Casini, Horacio and Hung, Ling-Yan and Myers, Robert C.",
    title = "{Relative Entropy and Holography}",
    eprint = "1305.3182",
    archivePrefix = "arXiv",
    primaryClass = "hep-th",
    doi = "10.1007/JHEP08(2013)060",
    journal = "JHEP",
    volume = "08",
    pages = "060",
    year = "2013"
}

@article{Dey:2014voa,
    author = "Dey, Anshuman and Mahapatra, Subhash and Sarkar, Tapobrata",
    title = "{Very General Holographic Superconductors and Entanglement Thermodynamics}",
    eprint = "1409.5309",
    archivePrefix = "arXiv",
    primaryClass = "hep-th",
    doi = "10.1007/JHEP12(2014)135",
    journal = "JHEP",
    volume = "12",
    pages = "135",
    year = "2014"
}

@article{Chakraborty:2014lfa,
    author = "Chakraborty, Somdeb and Dey, Parijat and Karar, Sourav and Roy, Shibaji",
    title = "{Entanglement thermodynamics for an excited state of Lifshitz system}",
    eprint = "1412.1276",
    archivePrefix = "arXiv",
    primaryClass = "hep-th",
    doi = "10.1007/JHEP04(2015)133",
    journal = "JHEP",
    volume = "04",
    pages = "133",
    year = "2015"
}

@article{Park:2015afa,
    author = "Park, Chanyong",
    title = "{Holographic entanglement entropy in the nonconformal medium}",
    eprint = "1501.02908",
    archivePrefix = "arXiv",
    primaryClass = "hep-th",
    doi = "10.1103/PhysRevD.91.126003",
    journal = "Phys. Rev. D",
    volume = "91",
    number = "12",
    pages = "126003",
    year = "2015"
}

@article{Mansoori:2015sit,
    author = "Mansoori, Seyed Ali Hosseini and Mirza, Behrouz and Darareh, Mahdi Davoudi and Janbaz, Sharooz",
    title = "{Entanglement Thermodynamics of the Generalized Charged BTZ Black Hole}",
    eprint = "1512.00096",
    archivePrefix = "arXiv",
    primaryClass = "gr-qc",
    doi = "10.1142/S0217751X16500676",
    journal = "Int. J. Mod. Phys. A",
    volume = "31",
    number = "12",
    pages = "1650067",
    year = "2016"
}

@article{Karar:2018ecr,
    author = "Karar, Sourav and Ghorai, Debabrata and Gangopadhyay, Sunandan",
    title = "{Holographic entanglement thermodynamics for higher dimensional charged black hole}",
    eprint = "1810.08037",
    archivePrefix = "arXiv",
    primaryClass = "hep-th",
    doi = "10.1016/j.nuclphysb.2018.11.018",
    journal = "Nucl. Phys. B",
    volume = "938",
    pages = "363--387",
    year = "2019"
}

@article{Allahbakhshi:2013rda,
    author = "Allahbakhshi, Davood and Alishahiha, Mohsen and Naseh, Ali",
    title = "{Entanglement Thermodynamics}",
    eprint = "1305.2728",
    archivePrefix = "arXiv",
    primaryClass = "hep-th",
    doi = "10.1007/JHEP08(2013)102",
    journal = "JHEP",
    volume = "08",
    pages = "102",
    year = "2013"
}

@article{Bhattacharya:2012mi,
    author = "Bhattacharya, Jyotirmoy and Nozaki, Masahiro and Takayanagi, Tadashi and Ugajin, Tomonori",
    title = "{Thermodynamical Property of Entanglement Entropy for Excited States}",
    eprint = "1212.1164",
    archivePrefix = "arXiv",
    primaryClass = "hep-th",
    reportNumber = "IPMU12-0220, YITP-12-99, IPMU12-0220; YITP-12-99",
    doi = "10.1103/PhysRevLett.110.091602",
    journal = "Phys. Rev. Lett.",
    volume = "110",
    number = "9",
    pages = "091602",
    year = "2013"
}

@article{Fischler:2012uv,
    author = "Fischler, Willy and Kundu, Arnab and Kundu, Sandipan",
    title = "{Holographic Mutual Information at Finite Temperature}",
    eprint = "1212.4764",
    archivePrefix = "arXiv",
    primaryClass = "hep-th",
    reportNumber = "UTTG-25-12, TCC-024-12, UTTG-25-12, TCC-024-12",
    doi = "10.1103/PhysRevD.87.126012",
    journal = "Phys. Rev. D",
    volume = "87",
    number = "12",
    pages = "126012",
    year = "2013"
}

@article{Fischler:2012ca,
    author = "Fischler, Willy and Kundu, Sandipan",
    title = "{Strongly Coupled Gauge Theories: High and Low Temperature Behavior of Non-local Observables}",
    eprint = "1212.2643",
    archivePrefix = "arXiv",
    primaryClass = "hep-th",
    doi = "10.1007/JHEP05(2013)098",
    journal = "JHEP",
    volume = "05",
    pages = "098",
    year = "2013"
}

@article{Hubeny:2012ry,
    author = "Hubeny, Veronika E.",
    title = "{Extremal surfaces as bulk probes in AdS/CFT}",
    eprint = "1203.1044",
    archivePrefix = "arXiv",
    primaryClass = "hep-th",
    reportNumber = "DCTP-12-15",
    doi = "10.1007/JHEP07(2012)093",
    journal = "JHEP",
    volume = "07",
    pages = "093",
    year = "2012"
}

@article{Solodukhin:2006xv,
    author = "Solodukhin, Sergey N.",
    title = "{Entanglement entropy of black holes and AdS/CFT correspondence}",
    eprint = "hep-th/0606205",
    archivePrefix = "arXiv",
    doi = "10.1103/PhysRevLett.97.201601",
    journal = "Phys. Rev. Lett.",
    volume = "97",
    pages = "201601",
    year = "2006"
}

@article{Cadoni:2010ztg,
    author = "Cadoni, Mariano and Melis, Maurizio",
    title = "{Holographic entanglement entropy of the BTZ black hole}",
    eprint = "0907.1559",
    archivePrefix = "arXiv",
    primaryClass = "hep-th",
    doi = "10.1007/s10701-010-9430-6",
    journal = "Found. Phys.",
    volume = "40",
    pages = "638--657",
    year = "2010"
}

@article{Maldacena:1997re,
    author = "Maldacena, Juan Martin",
    title = "{The Large $N$ limit of superconformal field theories and supergravity}",
    eprint = "hep-th/9711200",
    archivePrefix = "arXiv",
    reportNumber = "HUTP-97-A097, HUTP-98-A097",
    doi = "10.4310/ATMP.1998.v2.n2.a1",
    journal = "Adv. Theor. Math. Phys.",
    volume = "2",
    pages = "231--252",
    year = "1998"
}

@article{Witten:1998qj,
    author = "Witten, Edward",
    title = "{Anti de Sitter space and holography}",
    eprint = "hep-th/9802150",
    archivePrefix = "arXiv",
    reportNumber = "IASSNS-HEP-98-15",
    doi = "10.4310/ATMP.1998.v2.n2.a2",
    journal = "Adv. Theor. Math. Phys.",
    volume = "2",
    pages = "253--291",
    year = "1998"
}

@article{Gubser:1998bc,
    author = "Gubser, S. S. and Klebanov, Igor R. and Polyakov, Alexander M.",
    title = "{Gauge theory correlators from noncritical string theory}",
    eprint = "hep-th/9802109",
    archivePrefix = "arXiv",
    reportNumber = "PUPT-1767",
    doi = "10.1016/S0370-2693(98)00377-3",
    journal = "Phys. Lett. B",
    volume = "428",
    pages = "105--114",
    year = "1998"
}

@article{Ryu:2006bv,
    author = "Ryu, Shinsei and Takayanagi, Tadashi",
    title = "{Holographic derivation of entanglement entropy from AdS/CFT}",
    eprint = "hep-th/0603001",
    archivePrefix = "arXiv",
    reportNumber = "NSF-KITP-06-11, NSF-KITP-06-11",
    doi = "10.1103/PhysRevLett.96.181602",
    journal = "Phys. Rev. Lett.",
    volume = "96",
    pages = "181602",
    year = "2006"
}

@article{Hubeny:2007xt,
    author = "Hubeny, Veronika E. and Rangamani, Mukund and Takayanagi, Tadashi",
    title = "{A Covariant holographic entanglement entropy proposal}",
    eprint = "0705.0016",
    archivePrefix = "arXiv",
    primaryClass = "hep-th",
    reportNumber = "DCPT-07-13, KUNS-2069",
    doi = "10.1088/1126-6708/2007/07/062",
    journal = "JHEP",
    volume = "07",
    pages = "062",
    year = "2007"
}

@article{Vidal:2002zz,
    author = "Vidal, G. and Werner, R. F.",
    title = "{Computable measure of entanglement}",
    eprint = "quant-ph/0102117",
    archivePrefix = "arXiv",
    doi = "10.1103/PhysRevA.65.032314",
    journal = "Phys. Rev. A",
    volume = "65",
    pages = "032314",
    year = "2002"
}

@article{Plenio:2005cwa,
    author = "Plenio, M. B.",
    title = "{Logarithmic Negativity: A Full Entanglement Monotone That is not Convex}",
    eprint = "quant-ph/0505071",
    archivePrefix = "arXiv",
    doi = "10.1103/PhysRevLett.95.090503",
    journal = "Phys. Rev. Lett.",
    volume = "95",
    pages = "090503",
    year = "2005"
}

@article{Ishibashi:2011ws,
    author = "Ishibashi, Akihiro and Kodama, Hideo",
    title = "{Perturbations and Stability of Static Black Holes in Higher Dimensions}",
    eprint = "1103.6148",
    archivePrefix = "arXiv",
    primaryClass = "hep-th",
    doi = "10.1143/PTPS.189.165",
    journal = "Prog. Theor. Phys. Suppl.",
    volume = "189",
    pages = "165--209",
    year = "2011"
}

@article{Kodama:2003jz,
    author = "Kodama, Hideo and Ishibashi, Akihiro",
    title = "{A Master equation for gravitational perturbations of maximally symmetric black holes in higher dimensions}",
    eprint = "hep-th/0305147",
    archivePrefix = "arXiv",
    doi = "10.1143/PTP.110.701",
    journal = "Prog. Theor. Phys.",
    volume = "110",
    pages = "701--722",
    year = "2003"
}

@article{Ishibashi:2003ap,
    author = "Ishibashi, Akihiro and Kodama, Hideo",
    title = "{Stability of higher dimensional Schwarzschild black holes}",
    eprint = "hep-th/0305185",
    archivePrefix = "arXiv",
    doi = "10.1143/PTP.110.901",
    journal = "Prog. Theor. Phys.",
    volume = "110",
    pages = "901--919",
    year = "2003"
}

@article{Kodama:2003kk,
    author = "Kodama, Hideo and Ishibashi, Akihiro",
    title = "{Master equations for perturbations of generalized static black holes with charge in higher dimensions}",
    eprint = "hep-th/0308128",
    archivePrefix = "arXiv",
    doi = "10.1143/PTP.111.29",
    journal = "Prog. Theor. Phys.",
    volume = "111",
    pages = "29--73",
    year = "2004"
}

@article{Kodama:2007ph,
    author = "Kodama, Hideo",
    title = "{Perturbations and Stability of Higher-Dimensional Black Holes}",
    eprint = "0712.2703",
    archivePrefix = "arXiv",
    primaryClass = "hep-th",
    reportNumber = "KEK-COSMO-2",
    doi = "10.1007/978-3-540-88460-6_11",
    journal = "Lect. Notes Phys.",
    volume = "769",
    pages = "427--470",
    year = "2009"
}

@article{Chakrabortty:2022kvq,
    author = "Chakrabortty, Shankhadeep and Hoshino, Hironori and Pant, Sanjay and Sil, Karunava",
    title = "{A holographic study of the characteristics of chaos and correlation in the presence of backreaction}",
    eprint = "2206.12555",
    archivePrefix = "arXiv",
    primaryClass = "hep-th",
    doi = "10.1016/j.physletb.2023.137749",
    journal = "Phys. Lett. B",
    volume = "838",
    pages = "137749",
    year = "2023"}

@article{Chakrabortty:2020ptb,
    author = "Chakrabortty, Shankhadeep and Pant, Sanjay and Sil, Karunava",
    title = "{Effect of back reaction on entanglement and subregion volume complexity in strongly coupled plasma}",
    eprint = "2004.06991",
    archivePrefix = "arXiv",
    primaryClass = "hep-th",
    doi = "10.1007/JHEP06(2020)061",
    journal = "JHEP",
    volume = "06",
    pages = "061",
    year = "2020"
}

@article{Chakrabortty:2016xcb,
    author = "Chakrabortty, Shankhadeep and Dey, Tanay K.",
    title = "{Back reaction effects on the dynamics of heavy probes in heavy quark cloud}",
    eprint = "1602.04761",
    archivePrefix = "arXiv",
    primaryClass = "hep-th",
    doi = "10.1007/JHEP05(2016)094",
    journal = "JHEP",
    volume = "05",
    pages = "094",
    year = "2016"
}

@article{Chakrabortty:2011sp,
    author = "Chakrabortty, Shankhadeep",
    title = "{Dissipative force on an external quark in heavy quark cloud}",
    eprint = "1108.0165",
    archivePrefix = "arXiv",
    primaryClass = "hep-th",
    doi = "10.1016/j.physletb.2011.09.112",
    journal = "Phys. Lett. B",
    volume = "705",
    pages = "244--250",
    year = "2011"
}

@article{Pant:2024eno,
    author = "Pant, Sanjay and Parihar, Himanshu",
    title = "{Mixed state entanglement in deformed field theory at finite temperature}",
    eprint = "2412.19680",
    archivePrefix = "arXiv",
    primaryClass = "hep-th",
    doi = "10.1103/PhysRevD.111.086020",
    journal = "Phys. Rev. D",
    volume = "111",
    number = "8",
    pages = "086020",
    year = "2025"
}

@article{Karan:2023hfk,
    author = "Karan, Debanjan and Pant, Sanjay",
    title = "{Entanglement and Chaos near critical point in strongly coupled Gauge theory}",
    eprint = "2308.00018",
    archivePrefix = "arXiv",
    primaryClass = "hep-th",
    doi = "10.1140/epjc/s10052-024-12463-9",
    journal = "Eur. Phys. J. C",
    volume = "84",
    number = "2",
    pages = "113",
    year = "2024"
}

@article{Dutta:2019gen,
    author = "Dutta, Souvik and Faulkner, Thomas",
    title = "{A canonical purification for the entanglement wedge cross-section}",
    eprint = "1905.00577",
    archivePrefix = "arXiv",
    primaryClass = "hep-th",
    doi = "10.1007/JHEP03(2021)178",
    journal = "JHEP",
    volume = "03",
    pages = "178",
    year = "2021"
}

@article{Chaturvedi:2016kbk,
    author = "Chaturvedi, Pankaj and Malvimat, Vinay and Sengupta, Gautam",
    title = "{Entanglement thermodynamics for charged black holes}",
    eprint = "1601.00303",
    archivePrefix = "arXiv",
    primaryClass = "hep-th",
    doi = "10.1103/PhysRevD.94.066004",
    journal = "Phys. Rev. D",
    volume = "94",
    number = "6",
    pages = "066004",
    year = "2016"
}

@article{Calabrese:2012nk,
    author = "Calabrese, Pasquale and Cardy, John and Tonni, Erik",
    title = "{Entanglement negativity in extended systems: A field theoretical approach}",
    eprint = "1210.5359",
    archivePrefix = "arXiv",
    primaryClass = "cond-mat.stat-mech",
    doi = "10.1088/1742-5468/2013/02/P02008",
    journal = "J. Stat. Mech.",
    volume = "1302",
    pages = "P02008",
    year = "2013"
}

@article{Calabrese:2014yza,
    author = "Calabrese, Pasquale and Cardy, John and Tonni, Erik",
    title = "{Finite temperature entanglement negativity in conformal field theory}",
    eprint = "1408.3043",
    archivePrefix = "arXiv",
    primaryClass = "cond-mat.stat-mech",
    doi = "10.1088/1751-8113/48/1/015006",
    journal = "J. Phys. A",
    volume = "48",
    number = "1",
    pages = "015006",
    year = "2015"
}

@article{Calabrese:2013mi,
    author = "Calabrese, Pasquale and Tagliacozzo, Luca and Tonni, Erik",
    title = "{Entanglement negativity in the critical Ising chain}",
    eprint = "1302.1113",
    archivePrefix = "arXiv",
    primaryClass = "cond-mat.stat-mech",
    doi = "10.1088/1742-5468/2013/05/P05002",
    journal = "J. Stat. Mech.",
    volume = "1305",
    pages = "P05002",
    year = "2013"
}

@article{Wen:2015qwa,
    author = "Wen, Xueda and Chang, Po-Yao and Ryu, Shinsei",
    title = "{Entanglement negativity after a local quantum quench in conformal field theories}",
    eprint = "1501.00568",
    archivePrefix = "arXiv",
    primaryClass = "cond-mat.stat-mech",
    doi = "10.1103/PhysRevB.92.075109",
    journal = "Phys. Rev. B",
    volume = "92",
    number = "7",
    pages = "075109",
    year = "2015"
}

@article{Coser:2014gsa,
    author = "Coser, Andrea and Tonni, Erik and Calabrese, Pasquale",
    title = "{Entanglement negativity after a global quantum quench}",
    eprint = "1410.0900",
    archivePrefix = "arXiv",
    primaryClass = "cond-mat.stat-mech",
    doi = "10.1088/1742-5468/2014/12/P12017",
    journal = "J. Stat. Mech.",
    volume = "1412",
    number = "12",
    pages = "P12017",
    year = "2014"
}

@article{Rangamani:2014ywa,
    author = "Rangamani, Mukund and Rota, Massimo and Tonni, Erik",
    title = "{Entanglement negativity in holographic field theories}",
    eprint = "1412.6269",
    archivePrefix = "arXiv",
    primaryClass = "hep-th",
    doi = "10.1007/JHEP09(2014)003",
    journal = "JHEP",
    volume = "09",
    pages = "003",
    year = "2014"
}

@article{Jain:2023xta,
    author = "Jain, Parul and Pant, Sanjay and Parihar, Himanshu",
    title = "{Effect of backreaction on island, Page curve and mutual information}",
    eprint = "2311.08186",
    archivePrefix = "arXiv",
    primaryClass = "hep-th",
    doi = "10.1016/j.nuclphysb.2025.116991",
    journal = "Nucl. Phys. B",
    volume = "1018",
    pages = "116991",
    year = "2025"
}

@article{Kundu:2016dyk,
    author = "Kundu, Sandipan and Pedraza, Juan F.",
    title = "{Aspects of Holographic Entanglement at Finite Temperature and Chemical Potential}",
    eprint = "1602.07353",
    archivePrefix = "arXiv",
    primaryClass = "hep-th",
    doi = "10.1007/JHEP08(2016)177",
    journal = "JHEP",
    volume = "08",
    pages = "177",
    year = "2016"
}

@article{Nguyen:2017yqw,
    author = "Nguyen, Phuc and Devakul, Trithep and Halbasch, Matthew G. and Zaletel, Michael P. and Swingle, Brian",
    title = "{Entanglement of purification: from spin chains to holography}",
    eprint = "1709.07424",
    archivePrefix = "arXiv",
    primaryClass = "hep-th",
    doi = "10.1007/JHEP01(2018)098",
    journal = "JHEP",
    volume = "01",
    pages = "098",
    year = "2018"
}

\end{document}